\documentclass[prd,twocolumn,showkeys,preprintnumbers,superscriptaddress,nofootinbib,floatfix,longbibliography]{revtex4-2}

\usepackage{orcidlink}
\usepackage{lipsum}
\usepackage{amsmath}
\usepackage{amsfonts}
\usepackage{amssymb}
\usepackage{mathrsfs}
\usepackage{graphicx}
\usepackage{color}
\usepackage[dvipsnames]{xcolor}
\usepackage{bm}
\usepackage{blindtext}
\usepackage{wasysym}
\usepackage{hyperref}
\hypersetup{colorlinks=true,allcolors=blue}
\usepackage{booktabs}
\usepackage[normalem]{ulem}
\usepackage{fontawesome} 
\usepackage{multirow}
\usepackage[all]{hypcap}
\usepackage{lineno}

\begin{document}

\title{New constraints on non-unitary neutrino mixing from 8 years of IceCube DeepCore atmospheric neutrino data}

\author{Sharmistha Chattopadhyay\,\orcidlink{0009-0000-2435-4282}}
\email{sharmistha.c@iopb.res.in}
\affiliation{Institute of Physics, Sachivalaya Marg, Sainik School Post, Bhubaneswar 751005, India}
\affiliation{Homi Bhabha National Institute, Training School Complex, Anushakti Nagar, Mumbai 400094, India}

\author{Anil Kumar\,\orcidlink{0000-0002-8367-8401}}
\email{anil.k@iopb.res.in}
\affiliation{Institute of Physics, Sachivalaya Marg, Sainik School Post, Bhubaneswar 751005, India}

\author{Sanjib Kumar Agarwalla\,\orcidlink{0000-0002-9714-8866}}
\email{sanjib@iopb.res.in}
\affiliation{Institute of Physics, Sachivalaya Marg, Sainik School Post, Bhubaneswar 751005, India}
\affiliation{Homi Bhabha National Institute, Training School Complex, Anushakti Nagar, Mumbai 400094, India}

\preprint{IOP/BBSR/2026-10}

\begin{abstract}

The mixing between flavor and mass eigenstates of active neutrinos is described by a $3\times3$ unitary matrix. However, the presence of additional heavy sterile neutrino states can lead to a non-unitary neutrino mixing scenario. Atmospheric neutrinos, with their wide range of baselines and energies, provide an excellent probe of such effects. In particular, Earth matter effects in neutrino oscillations play an important role, as the neutral-current potential contributes non-trivially in the presence of non-unitarity. In this work, we use 8 years of publicly available atmospheric neutrino data of IceCube DeepCore to probe this non-unitary neutrino mixing scenario. This high-purity $\nu_\mu$ CC sample provides strong sensitivity, especially to the non-unitary parameters appearing at leading order in the $\nu_\mu \rightarrow \nu_\mu$ channel. The data sample is found to be consistent with the standard unitary mixing framework with no significant deviation. Using this data sample, we place the most stringent bound to date of $\alpha_{33} > -0.027$ at 90\% CL, while the other non-unitary parameters are constrained at competitive levels.

\end{abstract}

\maketitle

\section{Introduction and motivation}
\label{sec:intro}

\begin{figure*}[t]
	\centering
	\includegraphics[width=\textwidth]{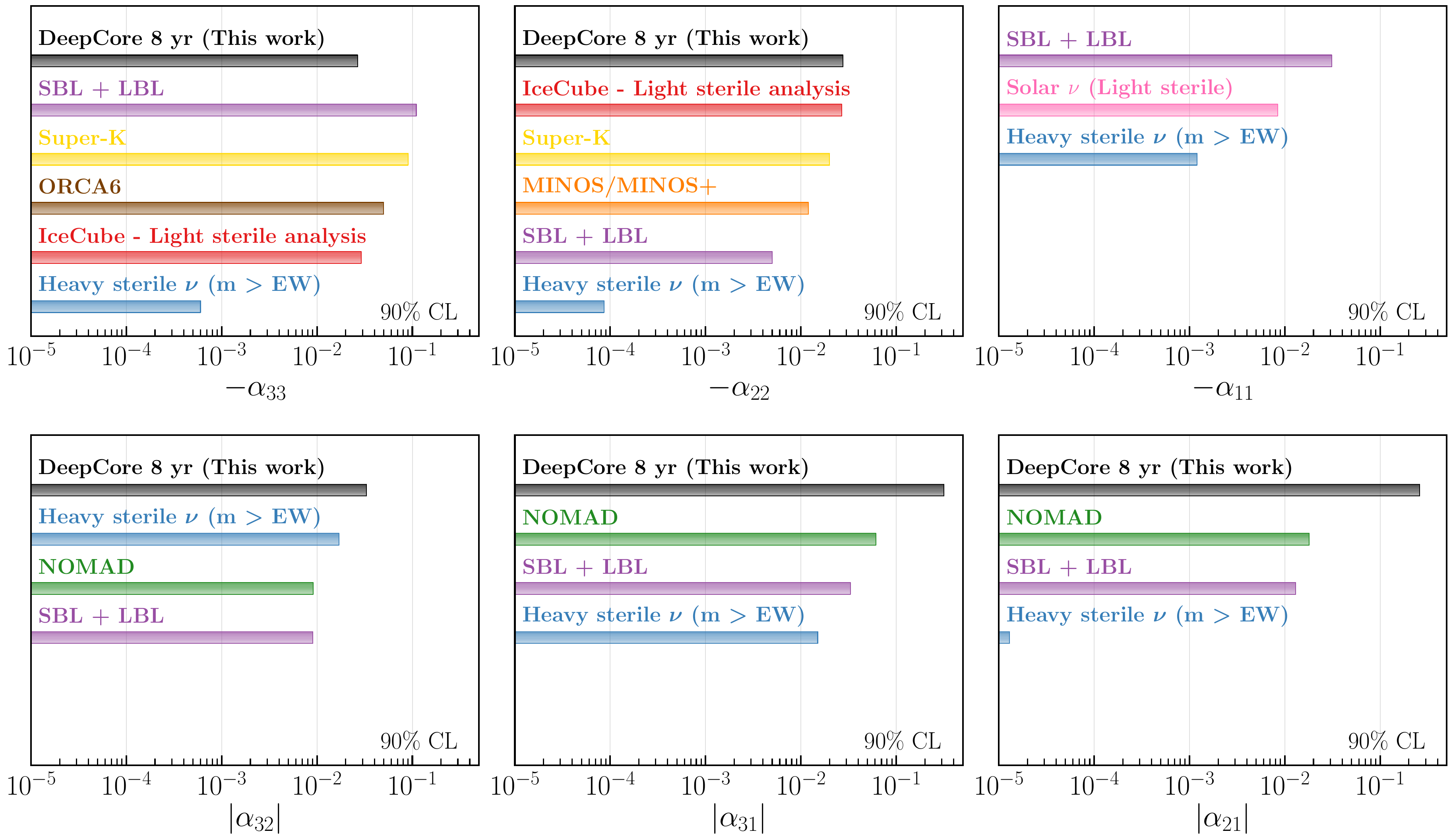}
	
	\caption{The constraints on the non-unitary neutrino mixing parameters at 90\% CL using 8 years of atmospheric neutrino data from IceCube DeepCore. Our results have been compared with the existing experimental limits, which include results from IceCube~\cite{IceCube:2024dlz}, Super-Kamiokande~\cite{Super-Kamiokande:2014ndf}, NOMAD~\cite{NOMAD:2003mqg,NOMAD:2001xxt}, KM3NeT/ORCA~\cite{KM3NeT:2025ftj}, MINOS~\cite{MINOS:2020iqj}, combined short- and long-baseline (SBL$+$LBL) analyses~\cite{Forero:2021azc}, solar data~\cite{Goldhagen:2021kxe}, and global fit to  flavor and electroweak precision observable data~\cite{Blennow:2023mqx}. For the heavy sterile neutrino scenario ($m>\mathrm{EW}$), the bounds on the diagonal parameters are taken from Ref.~\cite{Blennow:2025qgd}, while the bounds on the off-diagonal parameters are discussed in Appendix~\ref{app:heavy_sterile}. All remaining constraints, except those from (SBL$+$LBL) and ORCA6, are taken from Ref.~\cite{Blennow:2025qgd}. The bound quoted here for ORCA6 is without Feldman Cousins correction. Note that a negative sign has been placed in front of the diagonal parameters on the x-axis. This is because we only allow $\alpha_{ii}$ to take values less than zero in our parametrization in order to satisfy the equality $NN^\dagger + SS^\dagger = I$.}
	\label{fig:comparison}
\end{figure*}

The discovery of neutrino oscillations in 1998~\cite{Super-Kamiokande:1998kpq} by Super-Kamiokande opened a window to physics beyond the Standard Model (BSM) for the first time. Since then, over the past few decades, neutrino oscillation experiments spanning solar, atmospheric, reactor, and accelerator sectors have significantly enhanced our understanding of neutrino mixing. High-statistics data from various neutrino oscillation experiments have allowed precise measurements of the oscillation parameters. Most of the mixing parameters are now determined with percent-level accuracy from the global fits~\cite{NuFIT} to the available data. More recently, with only 59.1 days of data, Jiangmen Underground Neutrino Observatory (JUNO)~\cite{JUNO:2025gmd,Esteban:2026phq} has reached world-leading precision on the dominant oscillation parameters, measuring $\theta_{12}$ and $\Delta m_{21}^2$ with a relative $1\sigma$ precision at the percent level, consistent with the global fits. This marks an important step towards sub-percent precision in neutrino oscillation measurements. Additionally, next-generation long-baseline experiments such as DUNE~\cite{DUNE:2024wvj, DUNE:2021tad, DUNE:2020mra, DUNE:2020lwj, DUNE:2020txw, DUNE:2020ypp, DUNE:2020jqi, DUNE:2020fgq} are expected to provide precise measurements of $\delta_{CP}$, $\theta_{23}$, and $\Delta m_{31}^2$, and help resolve remaining degeneracies in the oscillation parameter space.

With the increasing precision of these measurements, neutrino oscillation experiments are not only sensitive to the standard parameters but can also probe possible subleading effects. This makes them a valuable tool to test the validity of the standard three-flavor neutrino oscillation framework and to search for new physics effects such as non-unitary neutrino mixing (NUNM). Independent constraints from collider experiments provide strong evidence for the existence of three active neutrino species. Precision measurements at the LEP collider have determined the number of light neutrino species to be $N_\nu = 2.9840 \pm 0.0082$, consistent with three active neutrinos~\cite{ALEPH:2005ab}. However, several anomalous results from short and long-baseline experiments~\cite{LSND:1995lje,LSND:2001aii,Kaether:2010ag,SAGE:2009eeu,Barinov:2022wfh} hint at the possibility of additional neutrino states. Such additional states can give rise to a scenario of non-unitarity, where the matrix corresponding to mixing among active states is a submatrix of a larger unitary matrix, implying the existence of more than three neutrino species. These additional neutrino states may be heavy and inactive, in the sense that they do not exhibit Standard Model (SM) interactions. Various BSM models (e.g., seesaw models)~\cite{Mohapatra:1986bd, Mohapatra:1979ia, Akhmedov:1995vm, Malinsky:2009gw, Minkowski:1977sc, Schechter:1980gr, Mohapatra:1980yp, Ma:1998dn}, which are often used to explain neutrino masses, commonly feature the existence of these additional neutrinos. If the new right-handed states are light enough to be kinematically accessible, they can be produced in the neutrino beam. This case is often referred to as the sterile neutrino scenario. Conversely, if the new right-handed states are too heavy to be produced, their existence may manifest itself as a deviation of the $3 \times 3$ mixing matrix of active states from unitarity. In the literature, these two mass regimes lead to different constraints on the non-unitary parameters.

In scenarios where additional heavy neutrino states extend the Standard Model Lagrangian, the full unitary matrix $U$ can be parameterized as
\begin{equation}
	U =
	\begin{pmatrix}
		N & S \\
		V & T
	\end{pmatrix}\,,
\end{equation}
where $U$ is an $n \times n$ matrix with 3 light and $n-3$ heavy neutrinos. Here, $N$ is the $3 \times 3$ mixing matrix between the light neutrinos, while $S$ and $V$ describe the active-heavy mixing, and $T$ describes the mixing among the heavy neutrinos. The matrix $U$ is unitary, with $UU^\dagger = 1$. However, since $N$ is a submatrix of $U$, it is not unitary, implying $NN^\dagger \neq 1$. The matrix $N$ is parameterized as
\begin{equation}
	N = (I + \hat{\alpha})U_{\text{PMNS}}\,.
	\label{eq:N_alpha_reln}	
\end{equation}
Here, $\hat{\alpha}$ is defined as
\begin{equation}
	\hat{\alpha} = 
	\begin{pmatrix}
		\alpha_{11} & 0 & 0 \\
		|\alpha_{21}|e^{i\phi_{21}} & \alpha_{22} & 0 \\
		|\alpha_{31}|e^{i\phi_{31}} & |\alpha_{32}|e^{i\phi_{32}} & \alpha_{33}
	\end{pmatrix}\,,
\end{equation}
where $\alpha_{ij}$ characterize small perturbations and $U_{\text{PMNS}}$ (also referred as the Pontecorvo-Maki-Nakagawa-Sakata matrix)~\cite{Pontecorvo:1957cp,Maki:1962mu,Pontecorvo:1967fh} is the standard $3\times3$ mixing matrix between the three active light neutrinos. The lower-triangular parameterization of $\hat{\alpha}$, first used by Okubo~\cite{Okubo:1962zzc}, provides a convenient description of non-unitarity in neutrino oscillations. This parameterization minimizes the impact of non-unitarity on the determination of the standard oscillation parameters and allows the effects of the non-unitary parameters to be treated separately. The diagonal elements of $\hat{\alpha}$ are real and are taken to be negative, such that $1+\alpha_{ii}$ is less than 1. This follows from the fact that $N$ is a submatrix of a larger unitary matrix and must satisfy the relation $NN^\dagger + SS^\dagger = I$, with
\begin{widetext}
	\begin{equation}
		NN^\dagger =
		\begin{pmatrix}
			(1+\alpha_{11})^2 & (1+\alpha_{11})\alpha_{21}^* & (1+\alpha_{11})\alpha_{31}^* \\
			(1+\alpha_{11})\alpha_{21} & (1+\alpha_{22})^2 + |\alpha_{21}|^2 & (1+\alpha_{22})\alpha_{32}^* + \alpha_{21}\alpha_{31}^* \\
			(1+\alpha_{11})\alpha_{31} & (1+\alpha_{22})\alpha_{32} + \alpha_{21}^*\alpha_{31} & (1+\alpha_{33})^2 + |\alpha_{31}|^2 + |\alpha_{32}|^2
		\end{pmatrix}.
		\label{eq:norm}
	\end{equation}
\end{widetext}
On the other hand, the off-diagonal elements can be complex and are related to the diagonal elements through the inequalities
\[
|\alpha_{ij}| \le \sqrt{(1-(1+\alpha_{ii})^2)(1-(1+\alpha_{jj})^2)} \, .
\]
In this work, we do not impose these inequalities; instead, we use the atmospheric neutrino data from IceCube DeepCore to constrain the non-unitary parameters.

The presence of such non-unitary effects can lead to observable deviations in neutrino oscillation probabilities. Consequently, it is important to constrain these parameters using experimental data. In recent years, several studies~\cite{Agarwalla:2021owd, KM3NeT:2025ftj, Soumya:2021dmy, Forero:2021azc, Blennow:2016jkn, Sahoo:2023mpj, Celestino-Ramirez:2023zox, Huang:2025znh, Antusch:2009pm, Escrihuela:2016ube, Rahaman:2021cgc, Trzeciak:2025hap, Kaur:2021rau, Blennow:2023mqx, Fernandez-Martinez:2016lgt, Declais:1994su, Super-Kamiokande:2014ndf, MINOS:2016viw, NOMAD:2003mqg, NOMAD:2001xxt} have been performed in the literature to constrain the non-unitary parameters. In this paper, we present a comprehensive analysis to search for the non-unitary neutrino mixing using the publicly available 8-year atmospheric neutrino data sample~\cite{DVN_B4RITM_2025} from IceCube DeepCore. This is a high-purity $\nu_\mu$ charged-current (CC) sample, where $\nu_\mu \rightarrow \nu_\mu$ is the dominant channel. This data sample is found to be consistent with the standard oscillation scenario with unitary mixing. Therefore, we use this data sample to place constraints on the NUNM parameters at 90\% CL as shown in Fig.~\ref{fig:comparison}. Our bounds are obtained by varying one NUNM parameter at a time while keeping the other NUNM parameters fixed to zero. The bottom panels present the constraints for the off-diagonal parameters, while the top panels show the constraints for the diagonal parameters. Note that the allowed physical parameter space for diagonal NUNM parameters corresponds to $\alpha_{ii} \le 0$; however, we have plotted $\alpha_{ii}$ on the x-axis with a negative sign so that the plotted values are positive and we can have a consistent representation similar to the off-diagonal parameters.

In Fig.~\ref{fig:comparison}, we also compare the existing bounds on the NUNM parameters from different experiments. As can be seen from the figure, the strongest constraints for most variables arise from electroweak precision observables ($\text{m} > \text{EW}$)~\cite{Blennow:2023mqx, Fernandez-Martinez:2016lgt}. In the presence of non-unitarity, the Standard Model couplings to the $Z$ and $W$ bosons are modified, which leads to strong limits on these parameters. The bounds from oscillation experiments, on the other hand, are comparatively weaker. Among neutrino oscillation experiments, constraint obtained using solar neutrino data~\cite{Goldhagen:2021kxe} provides the strongest bound on $\alpha_{11}$. However, in addition to the constraints shown in Fig.~\ref{fig:comparison}, further bounds can also be obtained on $\alpha_{11}$ from the recent limits placed on the $|U_{e4}|^2$ element\footnote{In the averaged-out sterile neutrino oscillation regime, limits on $|U_{\beta4}|^2$ can be translated into the corresponding bounds on $\alpha_{\beta\beta}$ using the relation $\alpha_{\beta\beta}\simeq \frac{1}{2}\lvert U_{\beta4}\rvert^2$~\cite{Blennow:2025qgd}.} from the light sterile analysis from Daya Bay and Bugey-3~\cite{MINOS:2020iqj} and RENO and NEOS data~\cite{RENO:2020hva}. For $\alpha_{21}$, $\alpha_{31}$, and $\alpha_{22}$, the strongest bounds come from a combined analysis of short-baseline experiments (such as NOMAD, NuTeV) and long-baseline experiments (such as MINOS/MINOS$+$, T2K, NO$\nu$A)~\cite{Forero:2021azc}. For $\alpha_{32}$, both NOMAD~\cite{NOMAD:2001xxt}, and the combined short and long-baseline analyses yield similar bounds. It can be seen that our results are competitive for the parameters $\alpha_{22}$ and $\alpha_{32}$, while providing the most stringent constraint on $\alpha_{33}$ till date. The strong sensitivities to these parameters arise from their dominant contribution to the $\nu_\mu \rightarrow \nu_\mu$ channel.

This paper is organized as follows. Section~\ref{sec:prob} discusses the formalism and the effect of non-unitarity on neutrino oscillation probabilities. Section~\ref{sec:gen_events} describes the IceCube DeepCore detector and the event sample. The analysis methodology and the results are presented in Sec.~\ref{sec:method} and Sec.~\ref{sec:results}, respectively. The summary and conclusion of the work can be found in Sec.~\ref{sec:summary}. In Appendix~\ref{app:heavy_sterile}, we present a discussion on the NUNM bounds for the heavy sterile neutrino.  Appendix~\ref{app:osc_prob} discusses the effect of non-unitary parameters on the absolute neutrino oscillation probabilities.  In Appendix \ref{app:bestfit}, we list our systematics parameters and their best-fit values.  A comparison of data and Monte Carlo (MC) is presented in Appendix~\ref{app:data_MC}.

\section{Formalism of NUNM}
\label{sec:prob}

This section presents the framework of non-unitary neutrino mixing and its impact on neutrino oscillation probabilities. In the first part, we describe how, in the presence of non-unitarity, the neutrino states are normalized. We then discuss the corresponding modification to the Hamiltonian, which governs neutrino propagation in matter. Finally, we use this framework to obtain the modified neutrino oscillation probabilities. In the second part, we discuss the effect of non-unitarity on oscillation probabilities and the different signatures shown by the various non-unitary parameters, along with the relevant channels for each parameter.

\subsection{NUNM Framework}

In the standard unitary case, the flavor ($|\nu_\alpha\rangle$) and mass ($|\nu_i\rangle$) states of active neutrinos are related by
\begin{equation}
	|\nu_\alpha \rangle = \sum_{i=1}^{3} U^*_{\alpha i}|\nu_i \rangle\,.
\end{equation}
However, in the presence of additional heavy neutrinos, this relation is modified. In such a scenario, the production and detection flavor states from charged-current interactions are related to the mass states as~\cite{Antusch:2009pm}
\begin{equation}
	|\nu_\alpha \rangle = \frac{1}{\sqrt{(NN^\dagger)_{\alpha \alpha}}}\sum_{i=1}^{3} N^*_{\alpha i}|\nu_i \rangle,
	\label{eq:states}
\end{equation}
where $\alpha$ stands for $e$, $\mu$ or $\tau$ flavor. Here, $1/\sqrt{(NN^\dagger)_{\alpha \alpha}}$ acts as a normalization factor for the neutrino states, ensuring that $\langle \nu_\alpha|\nu_\alpha \rangle = 1$ at $L=0$. However, we no longer have $\langle \nu_\alpha|\nu_\beta \rangle = 0$ at $L=0$, making the states non-orthogonal. This effect is commonly known as the zero-distance effect, {i.e.}, the appearance of a nonzero transition probability without neutrino propagation. 

In addition to the effects at production and detection, non-unitarity also modifies the propagation of neutrinos, especially in matter. The effective matter Hamiltonian in mass basis, governing the evolution of the neutrino mass eigenstates, can be written as
\begin{widetext}
	\begin{equation}
		H_m = 
		\begin{pmatrix}
			0 & 0 & 0 \\
			0 & \Delta m^2_{21} & 0 \\
			0 & 0 & \Delta m^2_{31}
		\end{pmatrix}
		+ N^\dagger
		\begin{pmatrix}
			V_{\rm CC} + V_{\rm NC} & 0 & 0\\
			0 & V_{\rm NC} & 0 \\
			0 & 0 & V_{\rm NC}
		\end{pmatrix}
		N,
		\label{eq:hamiltn}
	\end{equation}
\end{widetext}
where $V_{\rm CC}$ and $V_{\rm NC}$ are charged-current~\cite{Opher:1974drq,Langacker:1982ih} and neutral-current (NC) matter potentials, respectively, that are given as
\begin{equation}
	V_{\rm CC} = \pm\,\sqrt{2}G_F n_e = \pm\, 7.6 \times 10^{14} \times Y_e \left[\frac{\rho}{\rm g/cm^3}\right]\,{\rm eV}\,,
\end{equation}
and
\begin{equation}
	V_{\rm NC} = \mp\,\frac{1}{\sqrt{2}}G_Fn_n = \mp\, 3.8 \times 10^{14} \times Y_n \left[\frac{\rho}{\rm g/cm^3}\right]\,{\rm eV}\,.
\end{equation} 
Here, $G_F$ is the Fermi constant, and $n_e$ ($n_n$) is the electron (neutron) number density inside the Earth. $Y_e = n_e/(n_n + n_p)$ is the electron fraction for matter with mass density $\rho$, where $n_p$ denotes the proton number density inside Earth. The corresponding neutron fraction is given by $Y_n = n_n/(n_n + n_p)$. The positive (negative) and negative (positive) signs in $V_\text{CC}$ ($V_\text{NC}$) correspond to neutrinos and antineutrinos, respectively. In the standard unitary case, the term corresponding to neutral-current matter potential in the Hamiltonian is proportional to the identity matrix, and therefore, does not affect neutrino oscillation probabilities. However, as seen from Eq.~(\ref{eq:hamiltn}), in the case of non-unitarity, it becomes proportional to $N^\dagger N$ matrix, which is no longer unity. As a result, the neutral-current potential contributes non-trivially to the oscillation probability, enhancing the sensitivity to non-unitarity. Therefore, matter effects become particularly important while studying non-unitarity.

Atmospheric neutrinos are especially useful in this regard, as they propagate through different regions of the Earth and experience a broad range of baselines and matter densities. The matter effects experienced by these neutrinos lead to modifications in the effective mixing angles and mass-squared differences. In particular, the mixing angle $\theta_{13}$ undergoes resonant enhancement inside the Earth, reaching values as large as $45^\circ$ -- a phenomenon known as the Mikheyev-Smirnov-Wolfenstein (MSW) resonance~\cite{Wolfenstein:1977ue,Mikheyev:1985zog,Mikheev:1986wj}. Neutrinos passing through the mantle experience the MSW resonance in the energy range of 6 GeV to 10 GeV, which results in a significant modification to neutrino oscillation probabilities. Moreover, the large density jump at the core-mantle boundary results in an additional phenomenon known as the parametric resonance (PR)~\cite{Akhmedov:1998xq,Akhmedov:1998ui,Krastev:1989ix,Akhmedov:1988kd}, which also modifies the neutrino transition probability. Thus, baselines passing through the Earth's core experience strong matter effects due to the combined effect of propagation through both the mantle and the core, along with the higher matter density in the core. Hence, signatures of non-unitarity from such baselines are substantially enhanced due to the increased contribution from CC and NC matter potential terms.

To quantify the impact of non-unitary effects on atmospheric neutrino oscillations, we now derive the corresponding neutrino transition probability. To obtain the probability for neutrino oscillation from flavor $\alpha$ to $\beta$ in the presence of non-unitarity, we use Eq.~(\ref{eq:states}) with the effective Hamiltonian in Eq.~(\ref{eq:hamiltn}), and obtain the following expression,
\begin{equation}
	P_{\nu_\alpha \rightarrow \nu_\beta}(L)
	=
	\left| \langle \nu_\beta | \nu_\alpha(L) \rangle \right|^2
	=
	\frac{\left|\sum_i N_{\alpha i}^* e^{-iH_m L} N_{\beta i}\right|^2}
	{(NN^\dagger)_{\alpha \alpha}(NN^\dagger)_{\beta \beta}}\,,
	\label{eq:prob}
\end{equation}
where $(NN^\dagger)_{\alpha \alpha}$ and $(NN^\dagger)_{\beta \beta}$ account for the non-unitarity effects in the production and detection processes, respectively. It is interesting to observe that the normalization factors appearing in the denominator simplify considerably for parameters belonging to the $\tau$-row, namely $\alpha_{3i}$. For example, the probability for the $\nu_\mu \rightarrow \nu_\mu$ channel is given as
\begin{equation}
	P_{\nu_\mu \rightarrow \nu_\mu}(L)
	=
	\left| \langle \nu_\mu | \nu_\mu(L) \rangle \right|^2
	=
	\frac{\left|\sum_i N_{\mu i}^* e^{-iH_m L} N_{\mu i}\right|^2}
	{(NN^\dagger)_{\mu \mu}(NN^\dagger)_{\mu \mu}}\,.
\end{equation}
Looking at Eq.~(\ref{eq:norm}), we can see that when only one of the $\alpha_{3i}$ parameters is nonzero, with all other NUNM parameters set to zero, the matrix element $(NN^\dagger)_{\mu\mu}$ becomes unity. Thus, in this case, there will be no effect of normalization on the survival probability $P_{\nu_\mu \rightarrow \nu_\mu}$. However, note that the $\alpha_{3i}$ parameters can still affect $P_{\nu_\mu \rightarrow \nu_\mu}$ via matter effects during propagation. In the next subsection, we use Eq.~(\ref{eq:prob}) to discuss the impact of non-unitarity on neutrino oscillation probabilities. 

\subsection{Impact of NUNM on oscillation probability}

The presence of non-unitary neutrino mixing affects oscillation probabilities, particularly in matter, where the sensitivity is enhanced due to charged and neutral-current interactions. To account for the Earth's matter profile in our analysis, we consider a 12-layered PREM (Preliminary Earth Density Model) profile with electron fraction $Y_e$ equal to 0.4656 for the inner and outer core, and 0.4957 for the mantle (as discussed in Ref.~\cite{Rott:2015kwa}). If we assume the Earth to be neutral, then the corresponding neutron fraction is given by $Y_n = 1 - Y_e$. All oscillation probabilities presented in this work are numerically computed using Eq.~(\ref{eq:prob}) for three neutrino flavors in the presence of matter corresponding to the 12-layered PREM profile. To calculate oscillation probabilities in the presence of non-unitarity, we have implemented Eq.~(\ref{eq:prob}) in the publicly available software PISA~\cite{IceCube:2018ikn} provided by the IceCube Collaboration. 

\begin{figure*}[htp!]
	\centering
	\includegraphics[width=\linewidth]{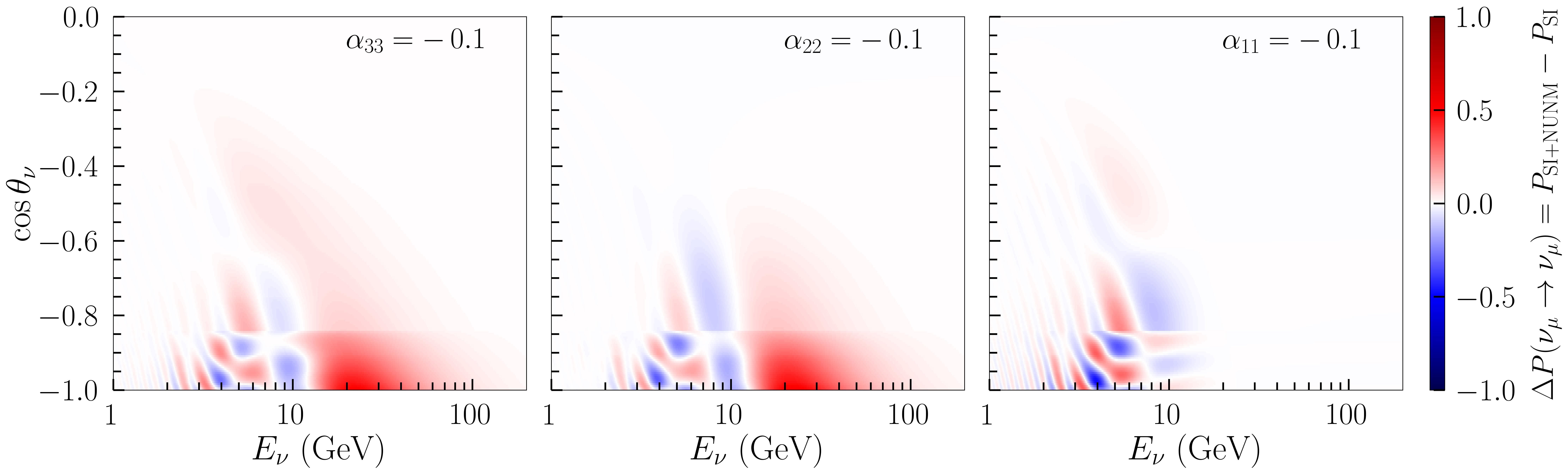}
	\caption{The difference of the three-flavor neutrino oscillation probability $P\,(\nu_\mu \rightarrow \nu_\mu)$ between the SI + NUNM and SI scenarios in the plane of $E_\nu$ and $\cos\theta_\nu$. The left, middle, and right panels show the difference for the diagonal NUNM parameters $\alpha_{33}$, $\alpha_{22}$, and $\alpha_{11}$, respectively. For these plots, we choose a representative value of $\alpha_{ii} = -\,0.1$ considering only one nonzero NUNM parameter at a time.}
	\label{fig:osc_diff_diag_neg}
\end{figure*}

Figure~\ref{fig:osc_diff_diag_neg} shows the difference in disappearance probability $P(\nu_\mu \rightarrow \nu_\mu)$ between the non-unitary and standard interaction (SI) scenarios for the diagonal parameters. The oscillogram differences are shown in the plane of neutrino energy ($E_\nu$) and cosine of zenith angle ($\cos \theta_\nu$) for the representative choice of $\alpha_{ii}=-\,0.1$. Parameters $\alpha_{33}$ and $\alpha_{22}$, shown in the left and middle panels of Fig.~\ref{fig:osc_diff_diag_neg}, respectively, primarily affect $\theta_{23}$ and modify the depth of the oscillation valley~\cite{Kumar:2020wgz, Kumar:2021lrn}. Moreover, we can see that these parameters exhibit a strong signal at lower energies and longer baselines, which can be attributed to the fact that these parameters enter at leading order in the $P(\nu_\mu \rightarrow \nu_\mu)$~\cite{Agarwalla:2021owd}. 

The right panel of Fig.~\ref{fig:osc_diff_diag_neg} shows the probability difference for the parameter $\alpha_{11}$, which couples to the standard charged-current matter potential, $V_\text{CC}$, and thus modifies the effective matter potential experienced by electron neutrinos. For example, in our case, where $\alpha_{11}<0$, this leads to a reduction in the effective matter potential, shifting the MSW resonance to higher zenith angles and energies. However, the signal for this parameter is much weaker compared to other diagonal parameters, as its effect is present only at the subleading order in the $\nu_\mu \rightarrow \nu_\mu$ channel. It enters at leading order only in the $\nu_e \rightarrow \nu_e$ channel~\cite{Agarwalla:2021owd}, which is subdominant in our data sample.

\begin{figure*}[htp!]
	\centering
	\includegraphics[width=\linewidth]{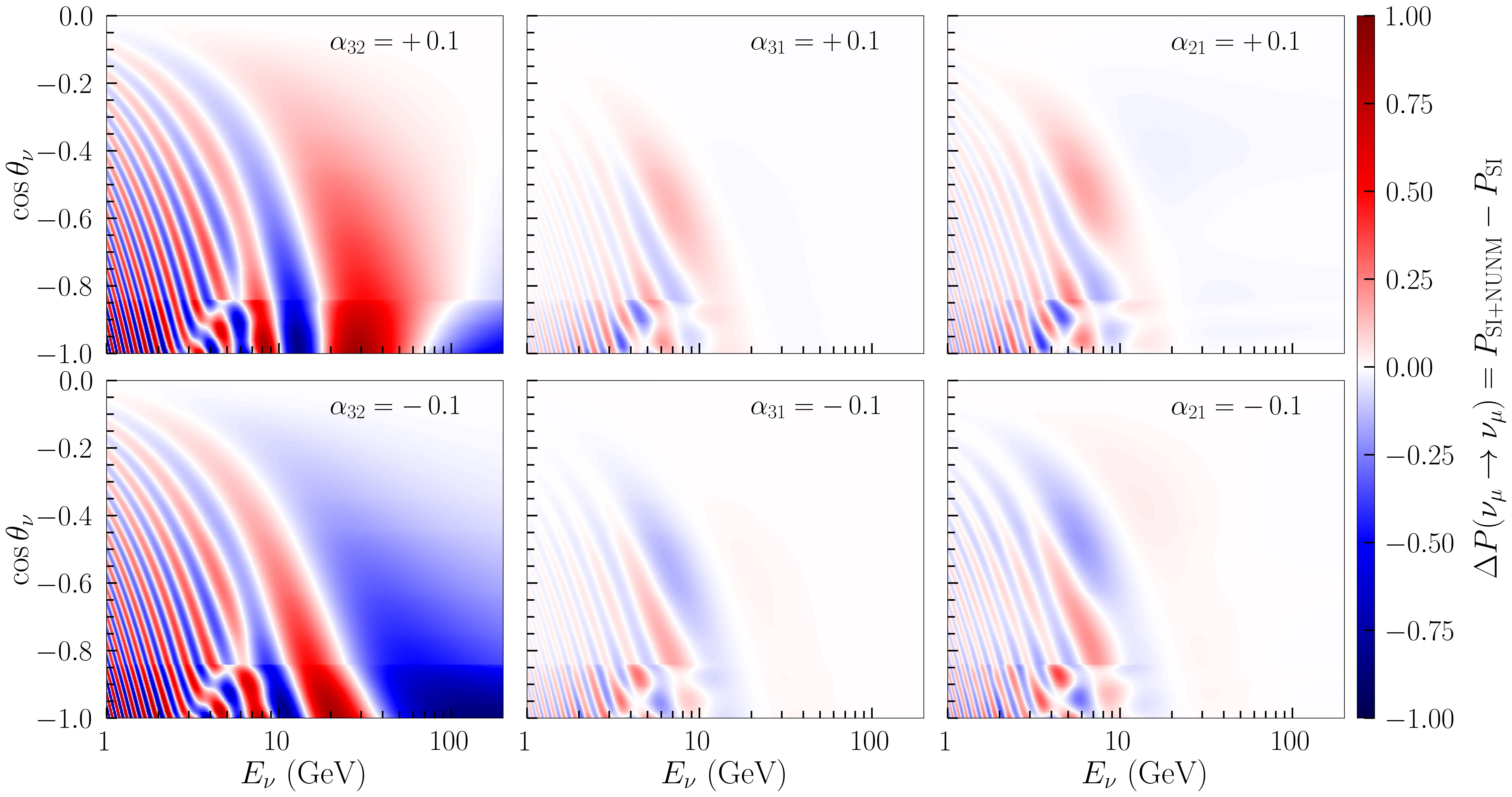}
	\caption{The difference of the three-flavor neutrino oscillation probability $P\,(\nu_\mu \rightarrow \nu_\mu)$ between the SI + NUNM and SI scenarios in the plane of $E_\nu$ and $\cos\theta_\nu$. The left, middle, and right panels show the difference for the off-diagonal NUNM parameters $\alpha_{32}$, $\alpha_{31}$, and $\alpha_{21}$, respectively. The top (bottom) panel shows the difference for a representative value of 0.1 ($-0.1$) for $\alpha_{ij}$ considering only one nonzero NUNM parameter at a time.}
	\label{fig:osc_diff_offdiag_mumu}
\end{figure*}

Figure~\ref{fig:osc_diff_offdiag_mumu} presents the difference in probability $P(\nu_\mu \rightarrow \nu_\mu)$ between the non-unitary and SI scenarios for the off-diagonal parameters. The top and bottom rows correspond to $\alpha_{ij} = +\,0.1$ and $\alpha_{ij} = -\,0.1$, respectively. The left, middle, and right panels show the oscillogram differences for $\alpha_{32}$, $\alpha_{31}$, and $\alpha_{21}$, respectively. Similar to the diagonal parameters, the signal regions for the off-diagonal parameters are also mainly concentrated at lower energies and larger baselines; although $\alpha_{32}$ shows a significant impact even at higher energies.

The probability difference for the off-diagonal parameter $\alpha_{32}$, shown in the left panel, demonstrates a strong impact in the $\nu_\mu \rightarrow \nu_\mu$ channel. This is due to the fact that $\alpha_{32}$ comes at a leading order in this channel~\cite{Agarwalla:2021owd}, resulting in pronounced signatures of non-unitarity. Moreover, the presence of a nonzero $\alpha_{32}$ gives rise to an interesting effect in the neutrino oscillation probability $-$ it causes the oscillation valley to bend. This can be understood as follows: in the standard scenario, $E_\nu \propto |\cos \theta_\nu|$; however, when $\alpha_{32}$ is nonzero, the relation modifies to~\cite{Sahoo:2023mpj}
\begin{equation}
	E_\nu \propto \frac{\Delta m^2_{32}|\cos \theta_\nu|}{(1 \pm 0.78\times\alpha_{32}.\rho.|\cos \theta_\nu|)},
\end{equation}
where $+$ ($-$) sign corresponds to neutrinos (antineutrinos), and $\rho$ denotes the line-averaged constant mass density along the propagation path. This relation is obtained by approximating $L \approx |2R\cos \theta_\nu|$ and assuming $Y_e = 0.5$. From this expression, it follows that when $\alpha_{32} > 0$, the denominator becomes greater than 1 for neutrinos, resulting in a decrease in energy for a given $\cos\theta_\nu$, and the valley bends inward. Similarly, when $\alpha_{32}<0$ for the case of neutrinos, the denominator decreases and the energy increases, resulting in the valley bending outward. For antineutrinos, the valley bends in the opposite direction. A more detailed discussion of this effect can be found in Ref.~\cite{Sahoo:2023mpj}.

The off-diagonal parameters $\alpha_{31}$ and $\alpha_{21}$, corresponding to the middle and right panels of Fig.~\ref{fig:osc_diff_offdiag_mumu}, result in the modification of the matter effects felt by the neutrinos. The positive (negative) values of $\alpha_{31}$ and $\alpha_{21}$ weaken (enhance) matter effects in the mantle region, leading to an increase (decrease) in the disappearance probability $P\,(\nu_\mu \rightarrow \nu_\mu)$. This effect can be seen in Fig.~\ref{fig:osc_offdiag_mumu} in Appendix~\ref{app:osc_prob}. However, the signal for $\alpha_{31}$ and $\alpha_{21}$ in the $\nu_\mu \rightarrow \nu_\mu$ channel is comparatively weaker than $\alpha_{32}$, as they appear at subleading orders in this channel, while entering at the leading order primarily in the $\nu_e \rightarrow \nu_\tau$ and $\nu_e \rightarrow \nu_\mu$ appearance channels. In the next section, we discuss the IceCube DeepCore detector and the event sample used in this study.

\section{Events at IceCube DeepCore}
\label{sec:gen_events}
This section discusses the event sample at IceCube DeepCore and the impact of NUNM on it. In the first subsection, we introduce the IceCube DeepCore detector and describe the data sample used in this work. We then discuss the impact of NUNM on the flux and cross section, and how the associated corrections can be incorporated. Finally, we present the effect of NUNM on the expected event distributions.

\subsection{IceCube DeepCore detector and data sample}

The IceCube Neutrino Observatory~\cite{IceCube:2016zyt} is a neutrino telescope located in the Antarctic ice of the South Pole. It is a Cherenkov detector, spanning 1 $\text{km}^2$ in area and 1 km in depth, making it a cubic-kilometer detector. The detector comprises 86 boreholes, each consisting of around 60 digital optical modules (DOMs), making a total of 5,160 DOMs. The DOMs are installed on vertical strings, between depths of 1450 m and 2450 m, in a hexagonal array with a horizontal spacing of 125 m and a vertical spacing of 17 m. This configuration enables the IceCube Neutrino Observatory to detect neutrinos with energies above approximately 100 GeV.

DeepCore~\cite{IceCube:2011ucd}, located at the bottom-center of IceCube, is a densely instrumented sub-detector featuring a shorter horizontal spacing of 42 m to 72 m and a vertical spacing of 7 m. It consists of DOMs with higher quantum efficiency photomultiplier tubes (PMTs) and is installed in the clearest part of the Antarctic ice, at depths between 2100 m and 2450 m. These features enable DeepCore to be sensitive to low-energy neutrinos, down to a few GeV, making it well-suited for studies of atmospheric neutrino oscillations.

Neutrinos interact in the ice to produce relativistic charged particles, which give rise to Cherenkov radiation during their propagation. The Cherenkov photons are detected by the DOMs, which produce electronic signals. The electronic waveforms collected by the detector are then subjected to various trigger conditions, followed by analog-to-digital conversion, after which the digitized signals are sent to the IceCube laboratory on the ice surface for further processing.
	
The pattern of energy deposit in the detector by secondary particles depends upon the neutrino flavor and the type of interaction. For example, the charged-current interaction of $\nu_\mu$ typically produces a muon that leaves a long track in the detector. In contrast, $\nu_e$ CC, $\nu_\tau$ CC, and neutral-current interactions of all flavors result in energy deposit in the form of cascades.   Note that some $\nu_\tau$ CC events can also yield muons from tau decay and therefore appear tracklike. Identification of neutrino flavors by distinguishing these event topologies is important for neutrino oscillation analysis.

The publicly available sample of IceCube DeepCore used in this work, also known as the \textit{golden event sample}, has been collected from 2011 to 2019, with a livetime of about 7.5 years~\cite{DVN_B4RITM_2025}. The golden event sample consists of only those photon hits that have undergone minimal scattering in ice. This sample, optimized for high-purity $\nu_\mu$ CC events, contains 21914 events in the reconstructed energy range from 6.3 GeV to 158.5 GeV. Note that both the data and the corresponding MC simulations of this sample are provided  by the IceCube Collaboration as part of a public data release~\cite{DVN_B4RITM_2025}. The sample has been processed by the IceCube Collaboration with improved detector calibrations, including updated in-situ single-photoelectron (SPE) charge calibration, optical detection efficiency calibration using in-situ data, and improved modeling of the ice properties. The sample also includes updated simulations of cosmic-ray and neutrino interactions, the propagation of outgoing charged particles and photons, and the detector response. Several filters are applied to reduce backgrounds from noise and cosmic muons, making the sample neutrino-dominated (as described in detail in Ref.~\cite{IceCubeCollaboration:2023wtb}).

The neutrino event observables, such as reconstructed energy ($E_{\mathrm{reco}}$) and direction ($\cos\theta_{\mathrm{reco}}$), are obtained using maximum-likelihood-based reconstruction methods. A boosted decision tree (BDT)~\cite{Friedman:2001wbq} is used to assign each event a particle identification (PID) score that quantifies the probability of an event being tracklike, with scores closer to zero corresponding to cascadelike events. Since neutrino and antineutrino events have similar detector signatures, they are not treated separately in this analysis. The data is binned in 10 logarithmic energy bins between 6.3 GeV and 158.5 GeV, and 10 linear bins in $\cos\theta_{\mathrm{reco}}$ from $-$1 to $0.1$. The final energy bin is taken twice as wide to maintain sufficient statistics. Events are further classified into mixed (PID: $0.55 - 0.75$) and tracklike (PID: $0.75 - 1.0$) categories. The cascadelike events with PID $<$ 0.55 are excluded by the IceCube Collaboration to ensure a high-purity $\nu_\mu$ CC sample.

\subsection{Impact of NUNM on flux and cross section}

Before discussing the effect of non-unitarity on the event rates, let us first understand how non-unitarity affects neutrino flux and cross section. If the analysis uses the flux and cross section that are obtained from theoretical calculations assuming SM, then corrections due to NUNM needs to be applied separately. In the presence of NUNM, flux from CC processes at production (taking into account one neutrino flavor) and cross section of CC and NC processes at  detection are modified to include correction factors as shown below~\cite{Kozynets:2024xgt},
\begin{subequations}
	\begin{align}
		\Phi_\alpha &= N_\alpha \, \Phi_\alpha^{\mathrm{SM}}, \\
		\sigma_\beta^{\mathrm{CC}} &= N_\beta \, \sigma_\beta^{\mathrm{SM, CC}}, \\
		\sigma_\beta^{\mathrm{NC}} &= N_\beta^2 \, \sigma_\beta^{\mathrm{SM, NC}},
	\end{align}
	\label{eq:flux_xsec}
\end{subequations}
where the normalization factors $N_\alpha$ and $N_\beta$ are $(NN^\dagger)_{\alpha \alpha}$ and $(NN^\dagger)_{\beta \beta}$, respectively. The flux $\Phi_\alpha^{\mathrm{SM}}$, and cross sections $\sigma_\beta^{\mathrm{SM, CC}}$ and $\sigma_\beta^{\mathrm{SM, NC}}$ are obtained from the theoretical calculation assuming SM. The number of events, given by the product of the flux, oscillation probability, and cross section, is then obtained by the following relation,
\begin{equation}
	n_{\text{events}} \sim \int dE \, \frac{d\Phi_\alpha(E)}{dE} \,
	P_{\nu_\alpha \to \nu_\beta}(E, L) \, \sigma_\beta(E) \, \epsilon(E) \,,
\end{equation}
where $\epsilon(E)$ is the detection efficiency. Now, we can see from the above relation, if the flux and cross section are taken from Eq.~(\ref{eq:flux_xsec}), then the  normalization factors in the denominator of the probability of Eq.~(\ref{eq:prob}) cancels with the correction factors in Eq.~(\ref{eq:flux_xsec}), and thus, no effect of non-unitarity is seen on the event rates.

In our analysis, we use the atmospheric neutrino flux model by Honda~\cite{Honda:2015fha} that has been derived using muon spectrometer data, and hence, will have effects of NUNM embedded in it. The cross-section model, on the other hand, has been simulated with GENIE using the SM framework, and does not include the NUNM correction factors mentioned in Eq.~(\ref{eq:flux_xsec}). Therefore, the NUNM correction factors associated with the cross section can be absorbed into the oscillation probability, leading to
\begin{align}
	\hat{P}_{\nu_\alpha \rightarrow \nu_\beta}(L)
	&=
	P_{\nu_\alpha \rightarrow \nu_\beta}(L) \times (NN^\dagger)_{\beta \beta} \nonumber \\
	&=
	\frac{\left|\sum_i N_{\alpha i}^* e^{-i H_m L} N_{\beta i}\right|^2}
	{(NN^\dagger)_{\alpha \alpha}}.
	\label{eq:prob_new_defn}
\end{align}
Note that the same correction factor can be applied to both CC and NC detection cross sections. This is because the relative normalization of NC to CC events, $\sigma_{NC}/\sigma_{CC}$, is already included as a free nuisance parameter in the analysis (see Sec.~\ref{sec:method}), which can account for the difference in correction factors associated with these two processes. In our analysis, we also have an additional nuisance parameter associated with an overall neutrino normalization, which can vary freely in the fit and account for the effects of the NUNM  correction in the cross section. Therefore, we do need to explicitly incorporate a factor for this NUNM correction, as shown in Eq.~(\ref{eq:prob_new_defn}), and instead, we use the probability from Eq.~(\ref{eq:prob}) in our analysis.

\subsection{Impact of NUNM on expected event distribution}

\begin{figure}[htbp]
	\centering
	\includegraphics[width=\linewidth]{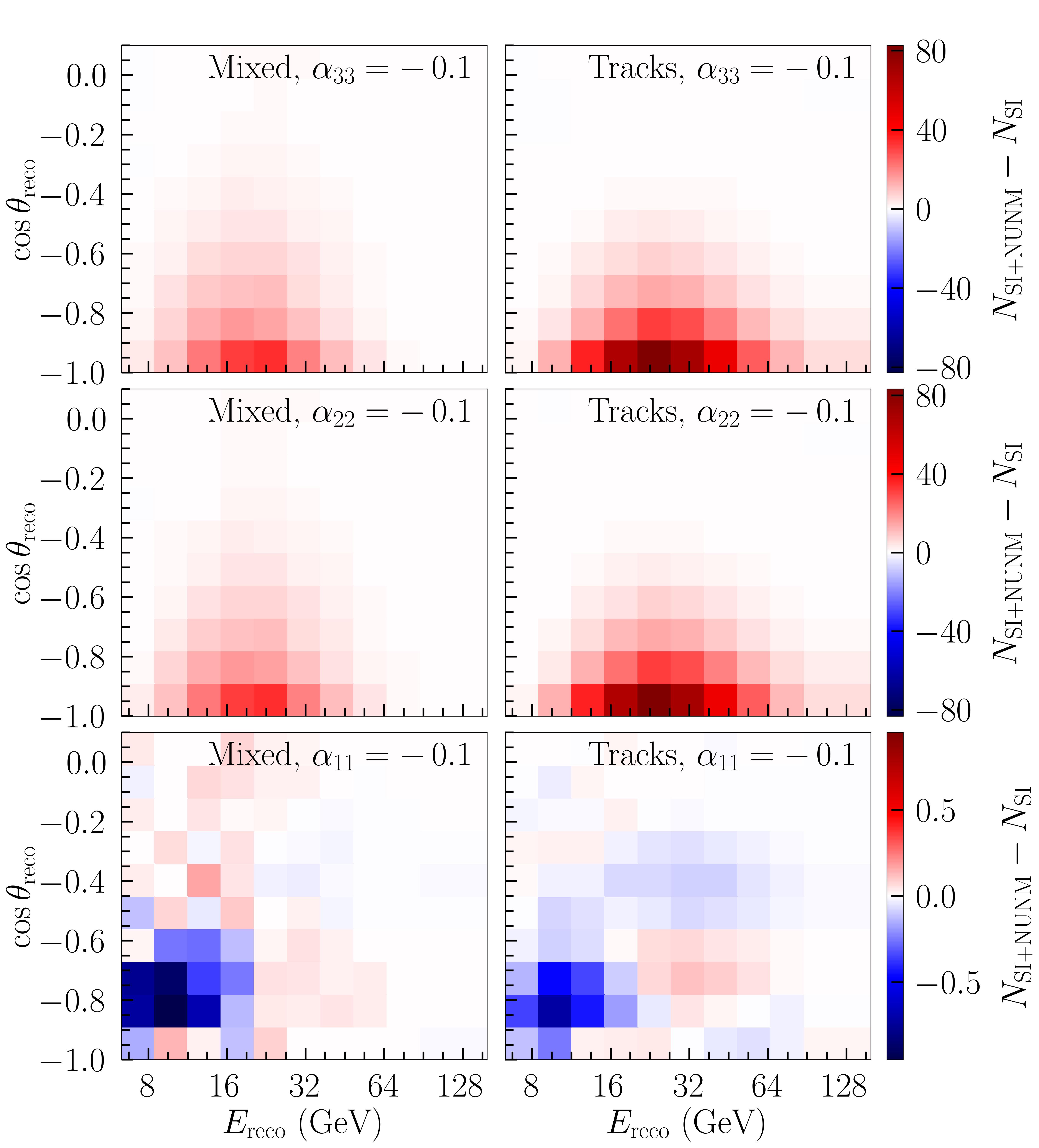}
	\caption{Expected event-difference distributions between the SI $+$ NUNM and SI scenarios plotted in the ($E_{\text{reco}}$, $\cos\theta_{\text{reco}}$) plane. The top, middle, and bottom panels correspond to the diagonal NUNM parameters $\alpha_{33}$, $\alpha_{22}$, and $\alpha_{11}$ with a representative value of $-0.1$. The left and right panels represent mixed and tracklike events, respectively.}
	\label{fig:event_diff_-0.1_diag}
\end{figure}

\begin{figure}[htbp]
	\centering
	\includegraphics[width=\linewidth]{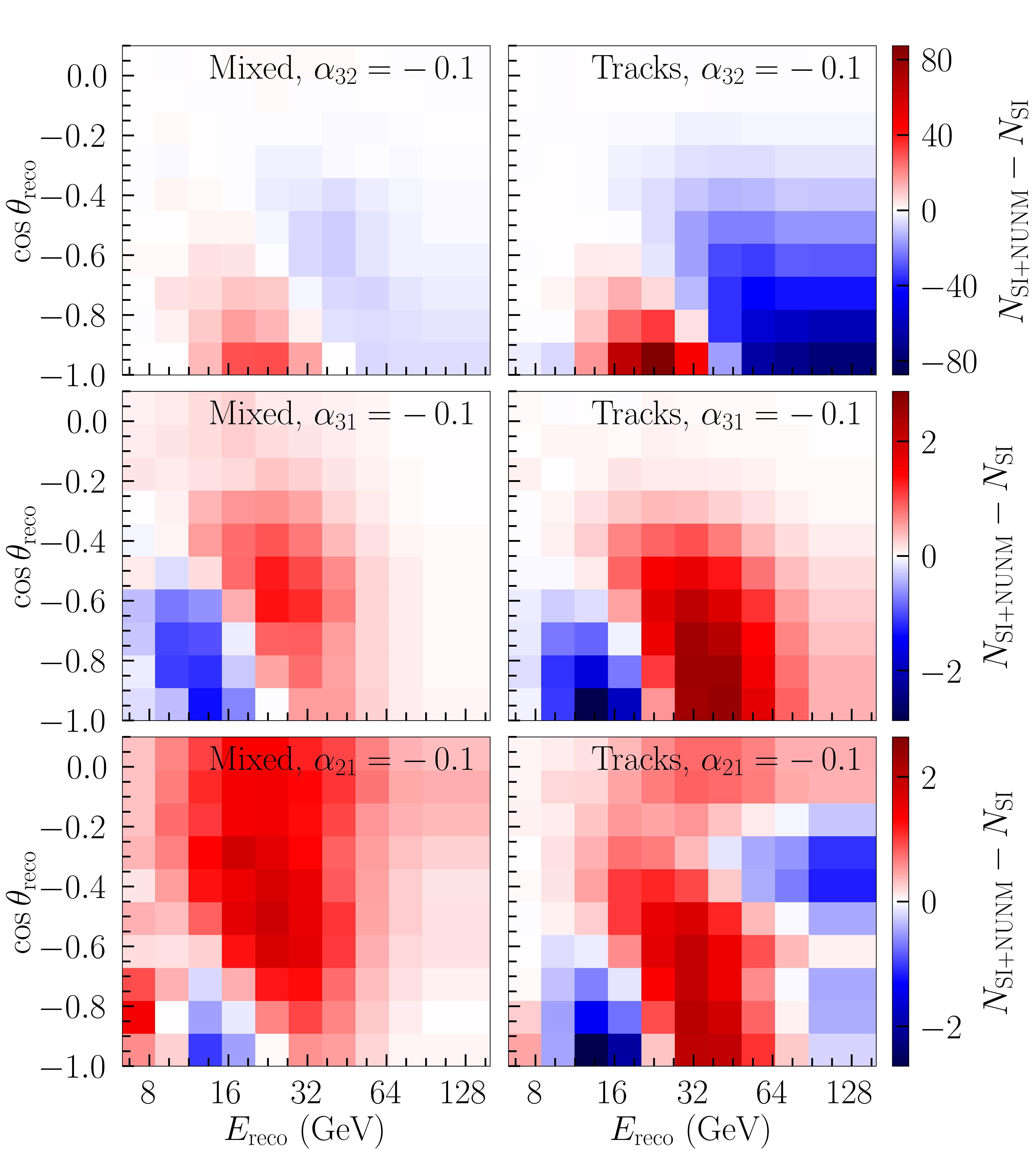}
	\caption{Expected event-difference distributions between the SI $+$ NUNM and SI scenarios plotted in the ($E_{\text{reco}}$, $\cos\theta_{\text{reco}}$) plane. The top, middle, and bottom panels correspond to the off-diagonal NUNM parameters $\alpha_{32}$, $\alpha_{31}$, and $\alpha_{21}$ with a representative value of $-\,0.1$. The left and right panels represent mixed and tracklike events, respectively.}
	\label{fig:event_diff_-0.1_offdiag}
\end{figure}

\begin{figure}[htbp]
	\centering
	\includegraphics[width=\linewidth]{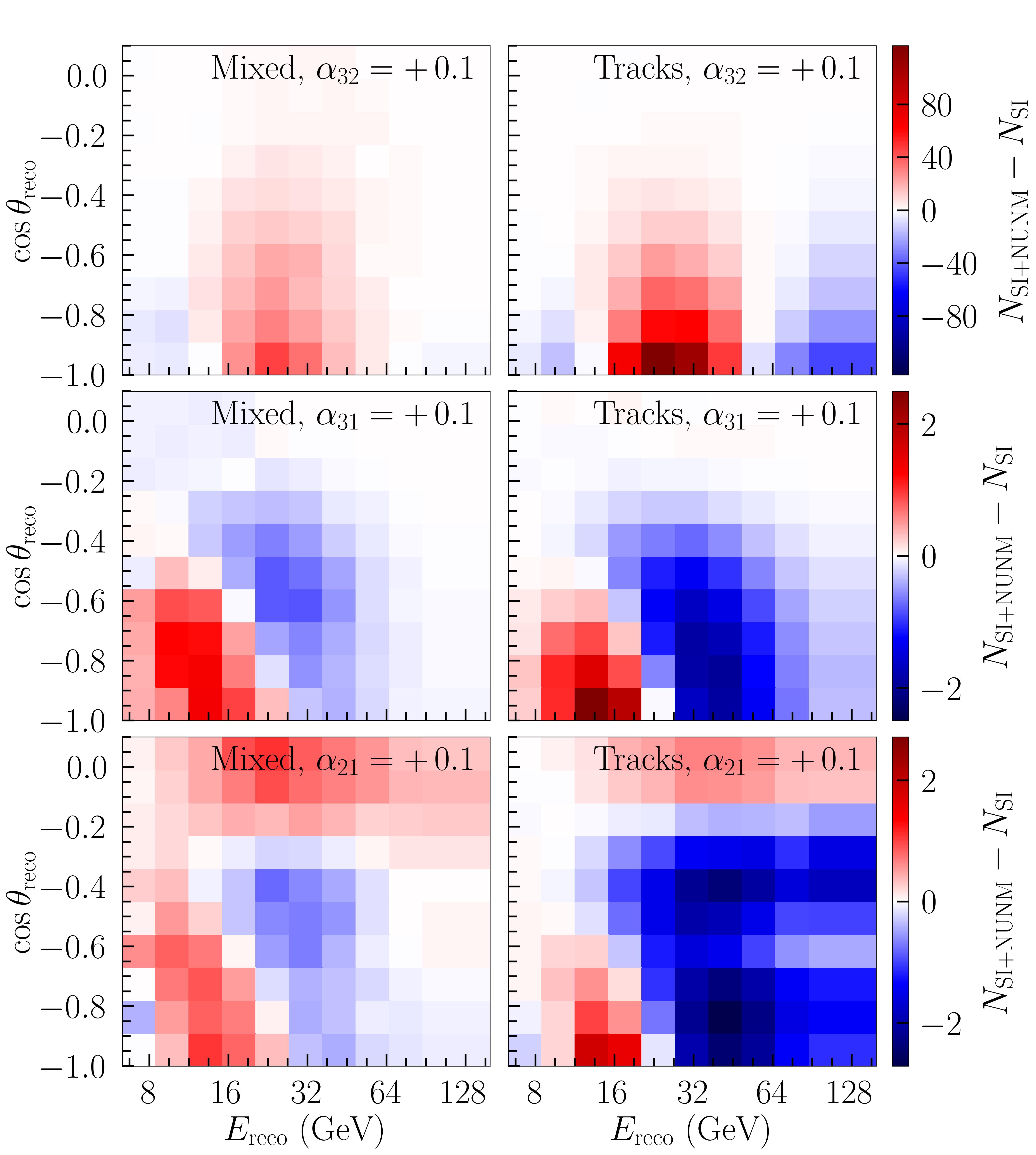}
	\caption{Expected event-difference distributions between the SI $+$ NUNM and SI scenarios plotted in the ($E_{\text{reco}}$, $\cos\theta_{\text{reco}}$) plane. The top, middle, and bottom panels correspond to the off-diagonal NUNM parameters $\alpha_{32}$, $\alpha_{31}$, and $\alpha_{21}$ with a representative value of $\,0.1$. The left and right panels represent mixed and tracklike events, respectively.}
	\label{fig:event_diff_0.1_offdiag}
\end{figure}

To illustrate the impact of NUNM on distribution of expected events at DeepCore, we plot the difference in event distributions between the SI $+$ NUNM and SI scenarios for the diagonal parameters in Fig.~\ref{fig:event_diff_-0.1_diag}, and for the off-diagonal parameters in Fig.~\ref{fig:event_diff_-0.1_offdiag} and Fig.~\ref{fig:event_diff_0.1_offdiag}, in the ($E_{\text{reco}}$, $\cos\theta_{\text{reco}}$) plane. The top, middle, and bottom panels in Fig.~\ref{fig:event_diff_-0.1_diag} correspond to the diagonal parameters $\alpha_{33}$, $\alpha_{22}$, and $\alpha_{11}$ with representative value of $-0.1$. Note that these event difference distributions are obtained by considering one NUNM parameter to be nonzero at a time. The left and right panels represent events with mixed and tracklike topologies, respectively. Consistent with the probability difference oscillograms, the signals for the diagonal NUNM parameters are predominantly concentrated at lower energies and longer baselines, with $\alpha_{33}$ and $\alpha_{22}$ showing larger event differences compared to $\alpha_{11}$, which is in agreement with their stronger impact on the disappearance channel. As discussed earlier, $\alpha_{33}$ and $\alpha_{22}$ modify the depth of the oscillation valley, which manifests as a change in the normalization of the event distributions in the signal region, thereby explaining the observed features in the figure.

Figures~\ref{fig:event_diff_-0.1_offdiag} and \ref{fig:event_diff_0.1_offdiag} show the event difference distributions for the off-diagonal parameters $\alpha_{32}$, $\alpha_{31}$, and $\alpha_{21}$ in the top, middle, and bottom panels, respectively. Figures~\ref{fig:event_diff_-0.1_offdiag} and \ref{fig:event_diff_0.1_offdiag} correspond to the representative values of $\alpha_{ij} = -\,0.1$ and $\alpha_{ij} = +\,0.1$, respectively. Similar to Fig.~\ref{fig:event_diff_-0.1_diag}, these figures highlight the regions in $\cos \theta_\text{reco}$ and $E_\text{reco}$ that are most sensitive to the off-diagonal NUNM parameters in our analysis. In agreement with the probability difference oscillograms, the off-diagonal parameters exhibit their impact primarily at lower energies and larger zenith angles, with $\alpha_{32}$ showing a noticeable effect even at higher energies. Moreover, as discussed in Sec.~\ref{sec:prob}, $\alpha_{32}$ has a stronger impact in the $\nu_\mu \rightarrow \nu_\mu$ channel compared to $\alpha_{31}$ and $\alpha_{21}$, which is reflected in the significantly larger event differences observed for $\alpha_{32}$. In the next section, we describe the analysis methodology followed in this work.

\section{Numerical Analysis}
\label{sec:method}

In this work, we adopt the Frequentist method for the statistical analysis. The analysis is based on a binned $\chi^2$, calculated between the expected and observed event counts, and is performed in bins of reconstructed energy, cosine of zenith angle, and PID. The test statistic used in this study is $\chi^2_\text{mod}$ (based on Ref.~\cite{IceCube:2017lak}), which is given by
\begin{equation}
	\chi^2_\text{mod} = \sum_{i \in \text{bins}} \frac{(N_i^{\text{exp}} - N_i^{\text{obs}})^2}
	{N_i^{\text{exp}} + (\sigma_i^{\text{sim}})^2}
	+
	\sum_{j \in \text{syst}}
	\frac{(s_j - \hat{s}_j)^2}{\sigma_{s_j}^2} \, ,
\end{equation}
where $N_i^{\text{exp}}$ ($N_i^{\text{obs}}$) denotes the expected (observed) number of events in the $i$th bin, and $\sigma_i^{\text{sim}}$ is the statistical uncertainty from simulation. The second term corresponds to the pull penalties accounting for external priors on nuisance parameters, where $s_j$, $\hat{s}_j$, and $\sigma_{s_j}$ of the $j$th nuisance parameter represent its test value, nominal value, and the associated systematic uncertainty, respectively. 

This analysis consists of 20 free systematic parameters, which are minimized in the fit. These include uncertainties associated with oscillation parameters, atmospheric flux, cross sections, detector systematics, and normalization. A detailed description regarding the nuisance parameters can be found in Ref.~\cite{IceCubeCollaboration:2023wtb}. For the oscillation parameters, we fix $\theta_{12}$, $\theta_{13}$, and $\Delta m_{21}^2$ to their values provided by NuFit v5.2, and set $\delta_{CP} = 0^\circ$ (as its impact is negligible), while keeping $\theta_{23}$ and $\Delta m_{31}^2$ free. For the atmospheric neutrino flux, we have seven systematic parameters, which account for uncertainties on the spectral index and production of mesons (as outlined in Ref.~\cite{Barr:2006it}). 

The nuisance parameters for cross-section  include four systematic uncertainties: axial mass for charged-current quasi-elastic interaction ($M_A^\text{CCQE}$), axial mass for charged-current resonant interaction ($M_A^\text{CCRES}$), Deep Inelastic Scattering (DIS), and the ratio of neutral-current to charged-current cross sections, $\sigma_\text{NC}/\sigma_\text{CC}$. Among these, $M_A^\text{CCQE}$ and $M_A^\text{CCRES}$ model the cross-section uncertainties associated with neutrino-nucleon interactions, while the DIS parameter effectively interpolates between the GENIE~\cite{GENIE-cross} and CSMS~\cite{Cooper-Sarkar:2011jtt} models describing Deep Inelastic Scattering.

To account for detector-related effects, we incorporate five systematic parameters. The DOM efficiency parameter characterizes the optical response of the digital optical modules. The parameters $p_0$ and $p_1$ describe the angular dependence of the DOM response, along with additional effects arising from variations in the optical properties of the re-frozen ice in the bore holes. Ice absorption and ice scattering parameters are included to account for photon absorption and scattering in the ice. Finally, two separate normalization parameters are included for neutrinos and cosmic muons. In addition to these systematic parameters, five physics parameters, namely the NUNM parameters $\alpha_{ij}$, are included in the fit, where only one NUNM parameter at a time is kept free while the other NUNM parameters are kept fixed to zero. Note that all computations are performed using the publicly available PISA~\cite{IceCube:2018ikn} software developed by the IceCube Collaboration. All results presented in this work assume a 12-layered PREM profile for the Earth, with $Y_e$ and $Y_n$ values as defined in Sec.~\ref{sec:prob}, and mass hierarchy as normal mass ordering (NO).

\section{Results}
\label{sec:results}

In this section, we present the constraints on the NUNM parameters obtained using the 8-year public data sample from the IceCube DeepCore detector. We first discuss the bounds on the diagonal parameters, followed by those on the off-diagonal parameters.

\subsection{Constraints on the diagonal NUNM parameters}

\begin{table}[t]
	\centering
	\begin{tabular}{lcc}
		\hline\hline
		Parameters & Best-fit values & $\Delta \chi^2_\text{SI-NUNM}$ \\
		\hline
		$\alpha_{22}$ & -0.00023 & 0.00012\\
		$\alpha_{33}$ & 0.00084 & 0.0017\\
		$\alpha_{21}$ & $|\alpha_{21}| = 0.126$, $|\phi_{21}| = 166.42^\circ$ & 0.74\\
		$\alpha_{31}$ & $|\alpha_{31}| = 0.197$, $|\phi_{31}| = 169.93^\circ$ & 1.37\\
		$\alpha_{32}$ & $|\alpha_{32}| = 0.0029$, $|\phi_{32}| = 4.46^\circ$ & 0.09\\
		
		\hline\hline
	\end{tabular}
	\caption{The best-fit values of the NUNM parameters and the associated $\Delta \chi^2$ to reject the SI scenario using the 8-year golden event sample of IceCube DeepCore.}
	\label{tab:nunm_params_i}
\end{table}

\begin{figure*}[htp!]
	\centering
	\includegraphics[width=\textwidth]{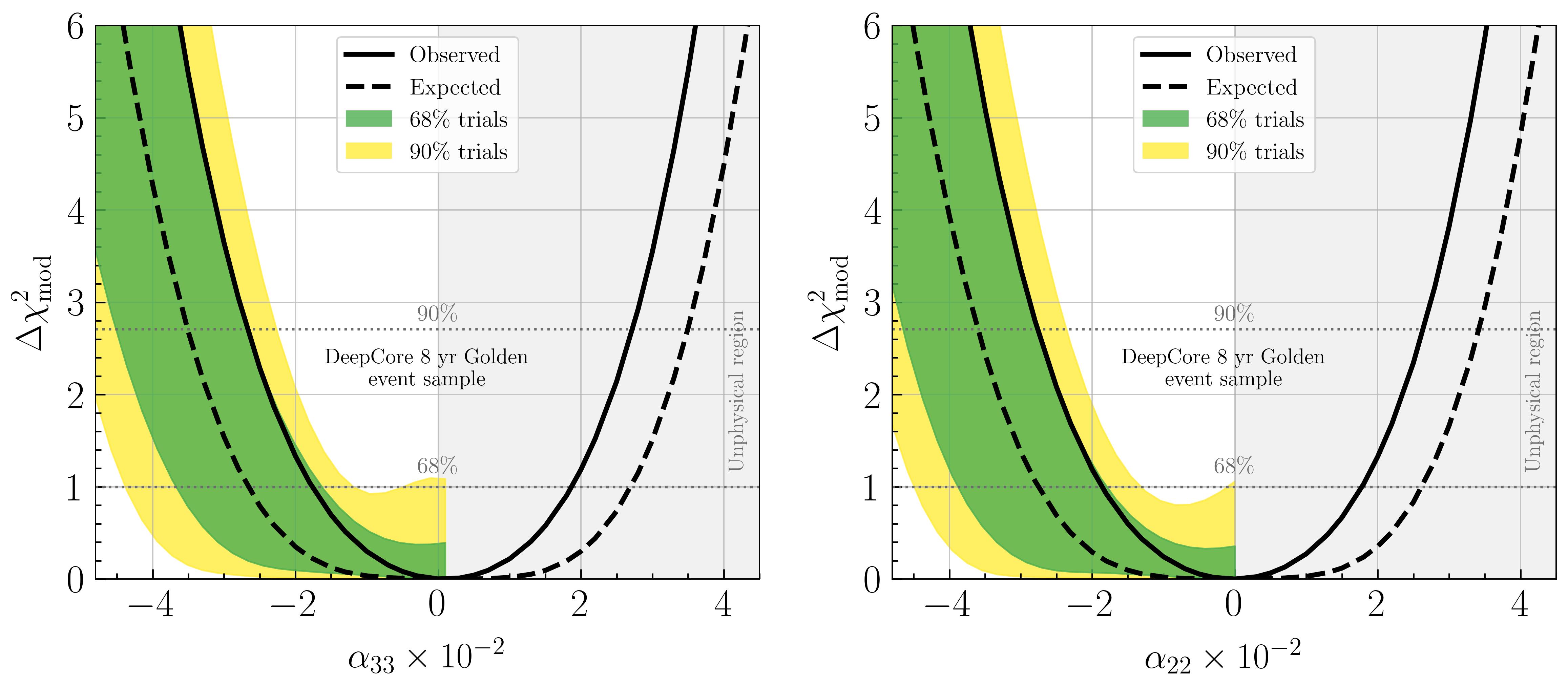}
	\caption{The solid and dashed black curves represent the observed and expected $\Delta \chi^2_{\rm mod}$ profile for the NUNM parameter $\alpha_{33}$ ($\alpha_{22}$) in the left (right) panel, respectively, where we use the 8-year golden event sample of IceCube DeepCore. In both panels, the 68\% and 90\% confidence levels for a single degree of freedom are indicated by horizontal dotted gray lines. The solid green (yellow) bands represent the 68\% (90\%) statistical uncertainties in $\Delta \chi^2_{\rm mod}$, which are obtained by fitting statistically fluctuated pseudoexperiments. The shaded gray region represents the unphysical region, corresponding to the parameter space with $\alpha_{ii}>0$, which does not satisfy the condition $NN^\dagger + SS^\dagger = I$.}
	\label{fig:a22_a33}
\end{figure*}

This section presents the search for the diagonal NUNM parameters $\alpha_{33}$ and $\alpha_{22}$, where we start by fitting the data with the NUNM model. The best-fit value of the parameter $\alpha_{33}$ is $8.45 \times 10^{-4}$ with a p-value of 0.25. The sensitivity to reject SI scenario can be quantified by $\Delta \chi^2_\text{SI-NUNM} = \chi^2_\text{SI} - \chi^2_\text{SI+NUNM}$, which is found to be 0.0017 for fit with $\alpha_{33}$. For the parameter $\alpha_{22}$, the best-fit is found to be at $- 2.28 \times 10^{-4}$ with a p-value of 0.25, and $\Delta \chi^2_\text{SI-NUNM}$ of 0.00012. The discussion on the data-MC agreement corresponding to different NUNM scenarios is given in Appendix~\ref{app:data_MC}. The best-fit values for all the NUNM parameters, along with their respective $\Delta \chi^2_\text{SI-NUNM}$, are summarized in Table~\ref{tab:nunm_params_i}. For completeness, the best-fit values of all the nuisance parameters obtained during the fit for each NUNM parameter are also summarized in Table~\ref{tab:combined_nuisance_parameters} of Appendix~\ref{app:bestfit}. For both $\alpha_{33}$ and $\alpha_{22}$, the observed best-fit values are found to be consistent with the standard unitary framework and do not show any significant deviation. Therefore, we use the DeepCore data sample to place constraints on these NUNM parameters.

To obtain constraints on the NUNM parameter, we scan it over a range of values, keeping it fixed in theory, and calculate the corresponding test statistic, $\Delta \chi^2_\text{mod}$, which is defined as
\begin{equation}
	\Delta \chi^2_\text{mod} = \chi^2_\text{mod}(\alpha_{ij} ~\text{fixed}) - \chi^2_\text{mod}(\alpha_{ij} ~\text{free}) \,.
\end{equation}
Here, $\chi^2_\text{mod}(\alpha_{ij} ~\text{fixed})$ and $\chi^2_\text{mod}(\alpha_{ij} ~\text{free})$ are calculated by fixing and freeing the physics parameter $\alpha_{ij}$, respectively, while minimization is performed over all the nuisance parameters. Figure~\ref{fig:a22_a33} shows the $\Delta \chi^2_\text{mod}$ profile for the parameters $\alpha_{33}$ (on the left) and $\alpha_{22}$ (on the right). The black solid curve in each plot denotes the observed $\Delta \chi^2_\text{mod}$ profile, obtained from fitting our model to the data. The black dashed curve in each plot denotes the expected sensitivity, obtained by fitting pseudoexperiment generated using the best-fit values of physics as well as nuisance parameters from the data fit. The dotted lines denote the confidence levels corresponding to 68\% and 90\% for 1 degree of freedom. The yellow (green) bands represent the 90\% (68\%) expected statistical uncertainties in $\Delta \chi^2_\text{mod}$, which are obtained by fitting the statistically fluctuated pseudoexperiments. Figure~\ref{fig:a22_a33} shows that the observed and the expected contours are in good agreement within the expected statistical bands.

\begin{table}[t]
	\centering
	\begin{tabular}{lc}
		\hline\hline
		Parameters & Bounds at 90\% CL \\
		\hline
		$\alpha_{22}$ & $\geq -0.0276$ \\
		$\alpha_{33}$ & $\geq -0.0266$ \\
		$|\alpha_{21}|$ & $\leq 0.259$ \\
		$\alpha_{21}$ & $[-0.259,\,0.152]$ \\
		$|\alpha_{31}|$ & $\leq 0.315$ \\
		$\alpha_{31}$ & $[-0.314,\, 0.201]$ \\
		$|\alpha_{32}|$ & $\leq 0.033$ \\
		$\alpha_{32}$ & $[-0.014,\, 0.017]$ \\
		\hline\hline
	\end{tabular}
	\caption{The observed constraints on the NUNM parameters at 90\% CL using the 8-year golden event sample of IceCube DeepCore.}
	\label{tab:nunm_params_ii}
\end{table}

\begin{figure*}[htp!]
	\centering
	\includegraphics[width=\textwidth]{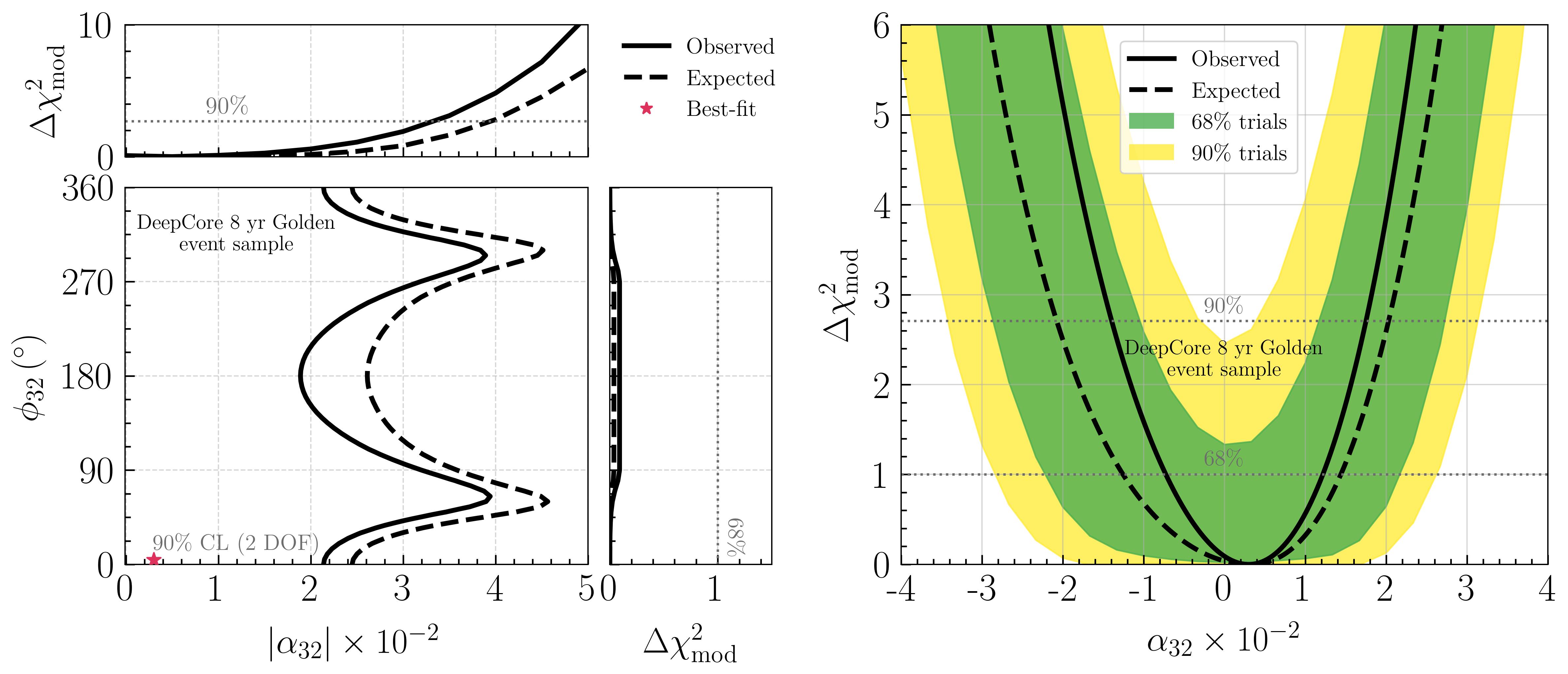}
	\caption{Left: the constraints on the magnitude $|\alpha_{32}|$ and the complex 	phase $\phi_{32}$ at 90\% CL (2 DOF) obtained from the 8-year golden event sample at IceCube DeepCore. The observed (expected) limits are represented by the black solid (black dashed) contour. The red star denotes the best-fit values for the magnitude and phase. The top (right) subpanel displays the one-dimensional projections for $|\alpha_{32}|$ ($\phi_{32}$), where the $\chi_\text{mod}^2$ is minimized with respect to $\phi_{32}$ ($|\alpha_{32}|$) and the nuisance parameters. The gray dotted lines in these subpanels denote the confidence levels for one degree of freedom. Right: the solid and dashed black curves represent the observed and expected profiles for $\alpha_{32}$ assuming it to be real. The solid green (yellow) band represent the 68\% (90\%) statistical uncertainties on $\Delta \chi^2_\text{mod}$. The 68\% and 90\% confidence levels for one degree of freedom are indicated by horizontal dotted gray lines.}
	\label{fig:a32}
\end{figure*}

The observed constraints on the NUNM parameter are determined from the points where the $\Delta \chi^2_\text{mod}$ curve (black solid curve) intersects the desired confidence-level thresholds for the corresponding degrees of freedom (DOF). At 90\% confidence level (CL), the lower limits on $\alpha_{33}$ and $\alpha_{22}$ obtained in this analysis are $-0.0266$ and $-0.0276$, respectively, which are also summarized in Table ~\ref{tab:nunm_params_ii}. As discussed previously, values of $\alpha_{ii} > 0$ are unphysical (shown by the shaded gray region); therefore, we quote only the lower bounds for the diagonal parameters and restrict the statistical bands to $\alpha_{ii} < 0$. However, for $\alpha_{33}$, the best-fit value lies on the side $\alpha_{ii} > 0$. Therefore, for this parameter, the band has been extended to the positive side. Our analysis exhibits strong sensitivity to the diagonal parameters $\alpha_{33}$ and $\alpha_{22}$, with the constraint obtained on $\alpha_{33}$ being the most stringent to date. This enhanced sensitivity arises from the leading-order dependence of the oscillation probability on $\alpha_{33}$ and $\alpha_{22}$ in the $\nu_\mu \rightarrow \nu_\mu$ channel, which dominates our event sample. We do not present any plot for $\alpha_{11}$, as we find its sensitivity to be extremely low. As discussed earlier, $\alpha_{11}$ is mainly constrained by the $\nu_e$ disappearance channel. However, as there is extremely low statistics for this channel, no meaningful constraint could be obtained on $\alpha_{11}$.

\subsection{Constraints on the off-diagonal NUNM parameters}

This section presents the search for the off-diagonal NUNM parameters $\alpha_{32}$, $\alpha_{31}$ and $\alpha_{21}$. Figure~\ref{fig:a32} shows the $\Delta\chi^2_{\text{mod}}$ profiles for $\alpha_{32}$. The left panel presents the constraints at 90\% CL (2 DOF) in the two-dimensional plane of the magnitude and phase of $\alpha_{32}$, along with the corresponding one-dimensional projections shown in the top and right subpanels. The best-fit values of the magnitude and phase of $\alpha_{32}$ (marked by a red star in the left panel) obtained from data fitting are $2.9 \times 10^{-3}$ and $4.457^\circ$, respectively, that are in good agreement with the standard interaction (SI) scenario, with a p-value of 0.24 and $\Delta \chi^2_\text{SI-NUNM}$ of 0.09. The solid (dashed) contour represents the observed result (expected sensitivity). The observed $\Delta\chi^2_{\text{mod}}$ contour is obtained by fitting the model to data, with the magnitude and phase fixed at each scan point in theory, while the minimization is performed over the nuisance parameters. The expected $\Delta\chi^2_{\text{mod}}$ contour is derived from a fit to a pseudoexperiment that is generated using the best-fit values from the corresponding data fit.
	
The top subpanel shows the one-dimensional projection of the magnitude obtained by minimizing over the phase, while the right subpanel shows the corresponding projection of the phase obtained by minimizing over the magnitude. The minimization over the nuisance parameters is performed in both subpanels. The dotted gray lines in the top and right subpanels indicate the 90\% and 68\% confidence levels, respectively.

\begin{figure*}[htp!]
	\centering
	\includegraphics[width=\textwidth]{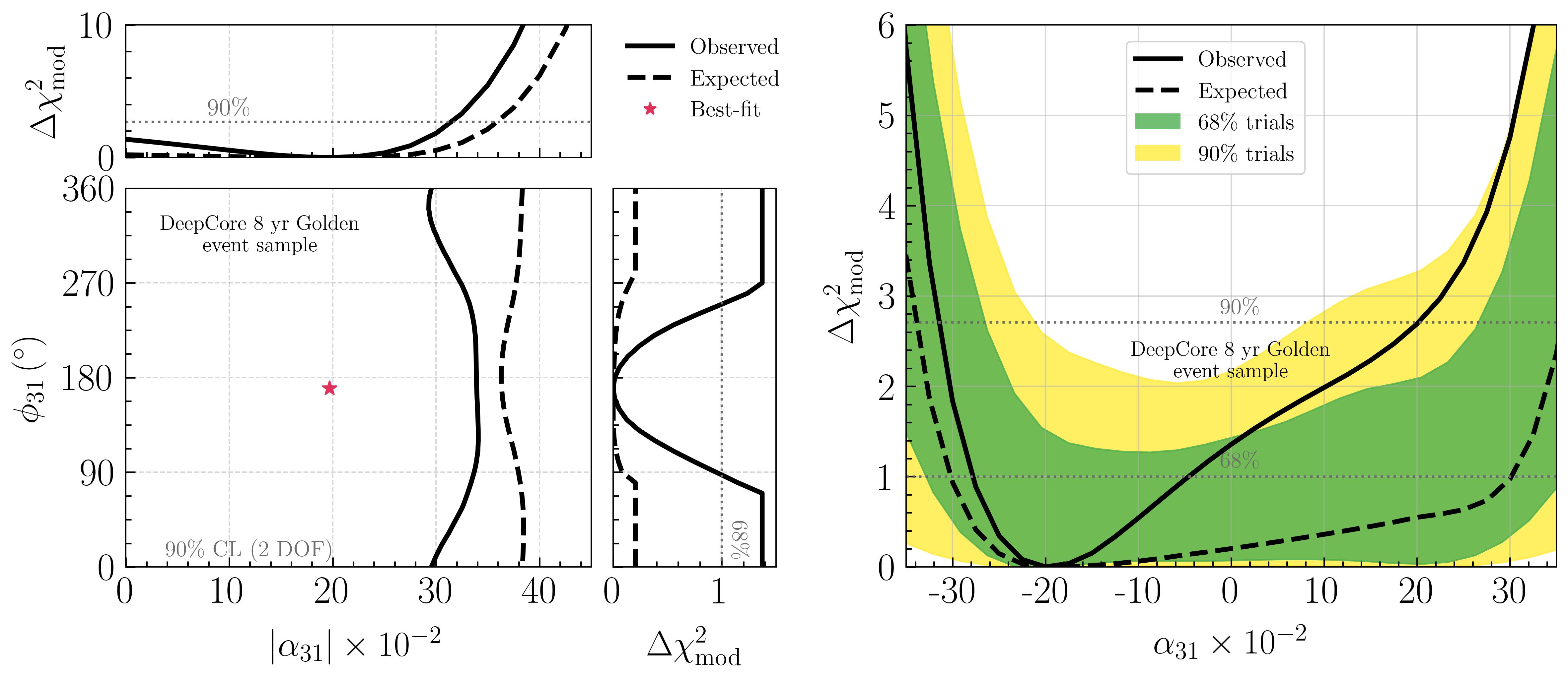}
	\caption{Left: the constraints on the magnitude $|\alpha_{31}|$ and the complex 	phase $\phi_{31}$ at 90\% CL (2 DOF) obtained from the 8-year golden event sample at IceCube DeepCore. The observed (expected) limits are represented by the black solid (black dashed) contour. The red star denotes the best-fit values for the magnitude and phase. The top (right) subpanel displays the one-dimensional projections for $|\alpha_{31}|$ ($\phi_{31}$), where the $\chi_\text{mod}^2$ is minimized with respect to $\phi_{31}$ ($|\alpha_{31}|$) and the nuisance parameters. The gray dotted lines in these subpanels denote the confidence levels for one degree of freedom. Right: the solid and dashed black curves represent the observed and expected profiles for $\alpha_{31}$ assuming it to be real. The solid green (yellow) band represent the 68\% (90\%) statistical uncertainties on $\Delta \chi^2_\text{mod}$. The 68\% and 90\% CL for one degree of freedom are indicated by horizontal dotted gray lines.}
	\label{fig:a31}
\end{figure*}

\begin{figure*}[htp!]
	\centering
	\includegraphics[width=\textwidth]{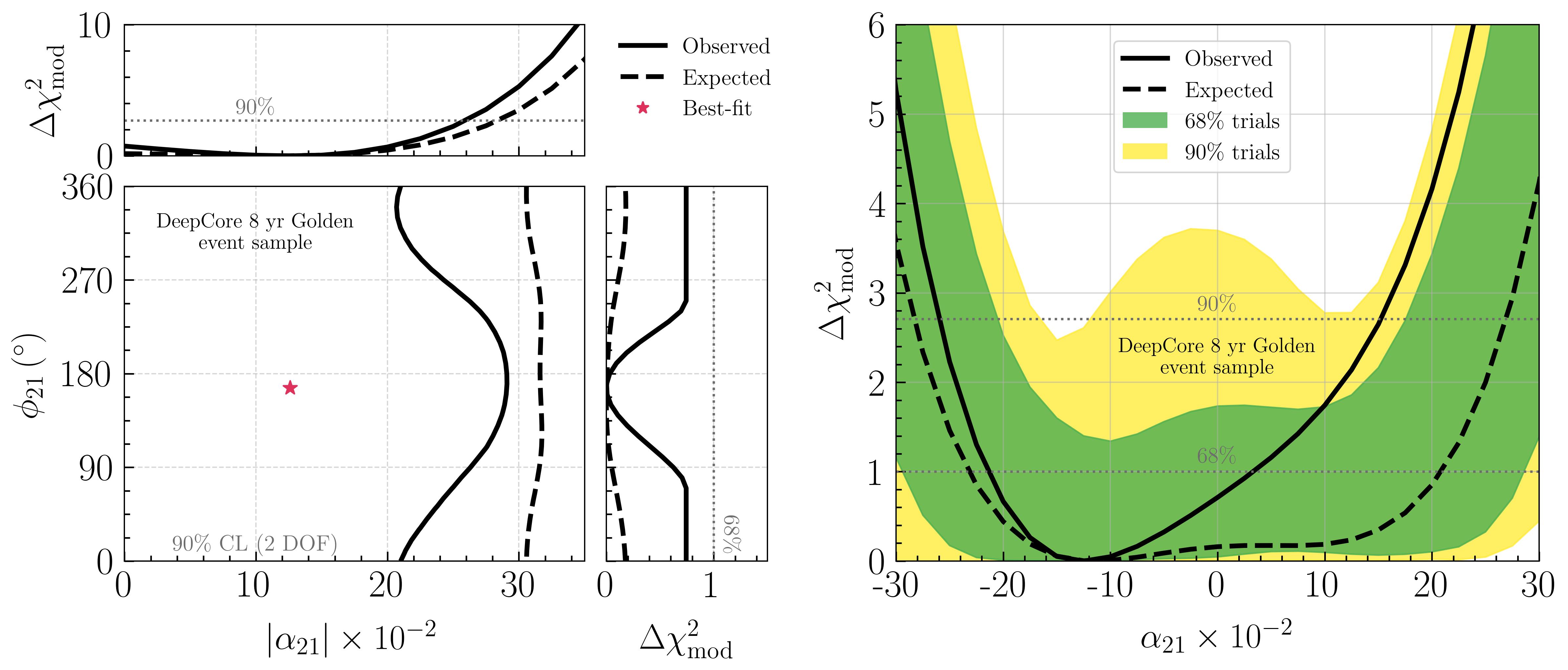}
	\caption{Left: the constraints on the magnitude $|\alpha_{21}|$ and the complex 	phase $\phi_{21}$ at 90\% CL (2 DOF) obtained from the 8-year golden event sample at IceCube DeepCore. The observed (expected) limits are represented by the black solid (black dashed) contour. The red star denotes the best-fit values for the magnitude and phase. The top (right) subpanel displays the one-dimensional projections for $|\alpha_{21}|$ ($\phi_{21}$), where the $\chi_\text{mod}^2$ is minimized with respect to $\phi_{21}$ ($|\alpha_{21}|$) and the nuisance parameters. The gray dotted lines in these subpanels denote the confidence levels for one degree of freedom. Right: the solid and dashed black curves represent the observed and expected profiles for $\alpha_{21}$ assuming it to be real. The solid green (yellow) band represent the 68\% (90\%) statistical uncertainties on $\Delta \chi^2_\text{mod}$. The 68\% and 90\% CL for one degree of freedom are indicated by horizontal dotted gray lines.}
	\label{fig:a21}
\end{figure*}

The two-dimensional contour exhibits weaker sensitivity around $\phi_{32} \approx 90^\circ$ and $270^\circ$, since $\alpha_{32}$ enters the $\nu_\mu$ disappearance probability as $|\alpha_{32}|\cos\phi_{32}$~\cite{Agarwalla:2021owd}, leading to reduced sensitivity at these values. From the one-dimensional projection, an upper bound of $|\alpha_{32}| \leq 0.033$ is obtained at 90\% confidence level. No meaningful constraint is obtained on the phase, as minimization over the magnitude drives the fitted value of magnitude close to zero, reducing the overall sensitivity.

The right panel of Fig.~\ref{fig:a32} shows the one-dimensional $\Delta \chi^2_\text{mod}$ profile for the same parameter $\alpha_{32}$, assuming it to be real. In this case, positive (negative) values $\alpha_{32}$ correspond to the phase fixed at $0^\circ$ ($180^\circ$). The solid (dashed) curve represents the observed (expected) $\Delta \chi^2_\text{mod}$ profile for $\alpha_{32}$. Using the observed curve, the constraints on $\alpha_{32}$ are found to be $[-0.014,\, 0.017]$ at 90\% confidence level. The green and yellow bands indicate the 68\% and 90\% expected statistical uncertainties, respectively. The observed and expected curves appear to be consistent within these bands.

Figure~\ref{fig:a31} presents the $\Delta \chi^2_\text{mod}$ profiles for the parameter $\alpha_{31}$. The left panel represents the allowed region at 90\% CL (2 DOF) in the two-dimensional plane of its magnitude and phase, along with their respective one-dimensional projections in the top and right subpanels. The solid (dashed) contours represent the observed (expected) $\Delta \chi^2_\text{mod}$ profiles. The best-fit values of the magnitude and phase of $\alpha_{31}$ are $1.97 \times 10^{-1}$ and $169.93^\circ$, respectively, which are shown by a red star in the two-dimensional plane. Thus, there is no significant deviation from the SI scenario, with a p-value of 0.26 and the sensitivity to reject the SI case, $\Delta \chi^2_\text{SI-NUNM}$, being 1.37. The upper bound on $\alpha_{31}$ obtained from the one-dimensional projection of the magnitude in the top subpanel is $|\alpha_{31}| \leq 0.315$. Similar to the previous case, no bound at 90\% CL could be obtained on the one-dimensional projection of the phase in the right subpanel. The right panel of Fig.~\ref{fig:a31} shows the one-dimensional $\Delta \chi^2_\text{mod}$ profile for $\alpha_{31}$ assuming it to be real, i.e., positive (negative) values corresponding to the phase fixed at $0^\circ$ ($180^\circ$). The observed constraints on $\alpha_{31}$ at 90\% CL is $[-0.314, 0.201]$. Moreover, the expected and the observed $\Delta \chi^2_\text{mod}$ profiles appear to be consistent with each other within the expected statistical bands.

Similarly, Fig.~\ref{fig:a21} represents the $\Delta \chi^2_\text{mod}$ profiles for the parameter $\alpha_{21}$. The left panel shows the constraints at 90\% CL (2 DOF) in the two-dimensional plane of magnitude and phase, with the top and right subpanels denoting the one-dimensional projection of magnitude and phase, respectively. The solid and dashed curves represent the observed and expected contours. The best-fit values of the magnitude and phase of $\alpha_{21}$ are $0.126$ and $166.42^\circ$, respectively, which are indicated by a red star in the two-dimensional plot. The p-value is 0.25, and the $\Delta \chi^2_\text{SI-NUNM}$ for rejecting the SI scenario is 0.74. For the one-dimensional projection on magnitude, the obtained upper bound at 90\% CL is $|\alpha_{21}| \leq 0.259$, while no meaningful constraint could be obtained from the one-dimensional projection of phase. The right panel of Fig.~\ref{fig:a21} demonstrates the one-dimensional $\Delta \chi^2_\text{mod}$ profile for $\alpha_{21}$ assuming it to be real. The constraints on $\alpha_{21}$ at 90\% CL from the right panel is $[-0.259, 0.152]$. Here also, the observed and the expected $\Delta \chi^2_\text{mod}$ profiles appear to be consistent with each other within the expected statistical bands.

As can be seen from our results, similar to the diagonal parameters, the best-fit values of the off-diagonal parameters are consistent with the standard unitary framework and show no significant deviation from it. Our analysis exhibits strong sensitivity to the parameter $\alpha_{32}$, and relatively weaker sensitivity for the parameters $\alpha_{31}$ and $\alpha_{21}$. This can be understood as follows: our event sample is dominated by $\nu_\mu$ CC events, with large statistics from the $\nu_\mu \rightarrow \nu_\mu$ channel. As discussed earlier, the parameter $\alpha_{32}$ appears at leading order in this channel and is therefore mainly constrained by it. In contrast, the parameters $\alpha_{31}$ and $\alpha_{21}$ are primarily constrained by the appearance channel $\nu_e \rightarrow \nu_\mu$, which has much lower statistics. As a result, $\alpha_{32}$ is more tightly constrained, while the bounds on $\alpha_{31}$ and $\alpha_{21}$ are comparatively weaker.

\section{Summary and conclusion}
\label{sec:summary}

With most of the oscillation parameters now determined at the percent level, neutrino oscillation experiments have become sensitive to physics scenarios beyond the standard three-flavor framework. A plethora of data from various neutrino oscillation experiments, spanning different energies and baselines, especially in the disappearance channels $\nu_\mu \rightarrow \nu_\mu$ and $\nu_e \rightarrow \nu_e$, and appearance channel $\nu_\mu \rightarrow \nu_e$, have allowed us to put strong constraints on the $e$ and $\mu$ rows of the PMNS matrix. However, the limited statistics of tau neutrinos have left the $\tau$ row less constrained, motivating the possibility of a non-unitary $3\times3$ neutrino mixing matrix.

In this work, we search for the non-unitary neutrino mixing using 8 years of publicly available atmospheric neutrino data from IceCube DeepCore. The data sample is observed to show no significant evidence for deviation of the mixing matrix from unitarity. The large statistics and high-purity $\nu_\mu$ CC sample enable us to place strong constraints on the non-unitary parameters, especially those appearing at leading order in the $\nu_\mu \rightarrow \nu_\mu$ channel. In particular, our study places the strongest bound on the parameter $\alpha_{33}$, while the constraints placed on $\alpha_{22}$ and $\alpha_{32}$ are competitive with existing bounds to date. However, constraints obtained on parameters $\alpha_{21}$ and $\alpha_{31}$ are comparatively weaker, as they appear only at subleading order in the $\nu_\mu \rightarrow \nu_\mu$ channel. 

With the recent deployment of the next-generation atmospheric neutrino detector IceCube Upgrade~\cite{Ishihara:2019aao,IceCubeCollaborationP:2025rpl}, which is now entering the data-taking phase, sensitivity to the non-unitary parameters is expected to increase. A larger data sample in the multi-GeV energy range and at long baselines, along with improved energy and direction resolution, will allow for a more stringent test of the unitarity of the neutrino mixing matrix. These advancements will further improve our understanding of neutrino mixing and help us to test the robustness of the standard three-flavor framework.

\section{Acknowledgments}

S.K.A., S.C., and A.K. acknowledge the support from the Department of Atomic Energy (DAE), Govt. of India. S.K.A and A.K receive support from the Swarnajayanti Fellowship (Sanction Order No. DST/SJF/PSA-05/2019-20) provided by the Department of Science and Technology (DST), Govt. of India, and the Research Grant (Sanction Order No. SB/SJF/2020-21/21) provided by the Anusandhan National Research Foundation (ANRF), Govt. of India, under the Swarnajayanti Fellowship. We warmly thank Christoph Ternes, Joao Coelho, Enrique Fernandez-Martinez, Daniel Naredo-Tuero, J Krishnamoorthi, Sadashiv Sahoo, and Sudipta Das for useful discussions. The numerical simulations are performed using the Dell PowerEdge R660 server at the Institute of Physics, Bhubaneswar, India.

\section{Data Availability}
 
In this work, we analyze the publicly available data from the IceCube Collaboration~\cite{DVN_B4RITM_2025}. Constraints on the NUNM parameters obtained in this study are available from the authors in the form of digitized files upon reasonable request.

\appendix

\setcounter{figure}{0}
\setcounter{table}{0}
\renewcommand{\thefigure}{\Alph{section}\arabic{figure}}
\renewcommand{\thetable}{\Alph{section}\arabic{table}}
\renewcommand{\theHfigure}{\Alph{section}\arabic{figure}}
\renewcommand{\theHtable}{\Alph{section}\arabic{table}}

\section{NUNM bounds for heavy sterile mixing}
\label{app:heavy_sterile}

In this appendix, we discuss the bounds on the off-diagonal NUNM parameters for the heavy sterile mixing scenario, where the sterile neutrino masses lie above the electroweak scale ($m > \mathrm{EW}$). For the parameter $\alpha_{21}$, a direct constraint can be obtained from the charged lepton flavor violating decay $\mu^+ \rightarrow e^+ \gamma$. In the presence of non-unitarity, the branching ratio of this process is related to the non-unitary parameter through~\cite{Blennow:2023mqx}
\begin{equation}
	\mathrm{BR}(\ell_\alpha \to \ell_\beta \gamma)
	\simeq
	\frac{3\alpha}{2\pi}
	\left| \eta_{\alpha\beta} \right|^2 ,
\end{equation}
where $\alpha$ is the fine-structure constant and $\eta_{\alpha\beta}=\alpha_{\alpha\beta}/2$. The most recent upper limit on the branching ratio, reported by the MEG II experiment~\cite{MEGII:2025gzr}, is $\mathrm{BR}(\mu^+ \rightarrow e^+ \gamma) < 1.5 \times 10^{-13}$ at 90\% CL. Using this value in the above equation, we obtain the bound $|\alpha_{21}| < 1.3 \times 10^{-5}$, which corresponds to the bound shown in Fig.~\ref{fig:comparison}.

For the parameters $\alpha_{31}$ and $\alpha_{32}$, we use the limits on $\eta_{\tau e}$ and $\eta_{\tau\mu}$ quoted in Table~6 of Ref.~\cite{Blennow:2023mqx}. The bounds reported in Ref.~\cite{Blennow:2023mqx} are given at 95\% CL ($\sim\,2\sigma$) and are derived from charged lepton flavor-violating processes. Since the branching ratio depends quadratically on $\eta$, we rescale the quoted limits by a factor of $\sqrt{1.645/2}$ to obtain the corresponding bounds at 90\% CL ($\sim\,1.645\sigma$). The corresponding bounds on $\alpha_{31}$ and $\alpha_{32}$ are then obtained using the relation $\alpha_{\alpha\beta}=2\eta_{\alpha\beta}$. The resulting limits at 90\% CL are $\alpha_{31} < 1.5 \times 10^{-2}$ and $\alpha_{32} < 1.7 \times 10^{-2}$, which are shown in Fig.~\ref{fig:comparison}.

\section{Effect of the NUNM parameters on neutrino oscillation probabilities}
\label{app:osc_prob}

\begin{figure*}[htp!]
	\centering
	\includegraphics[width=\textwidth]{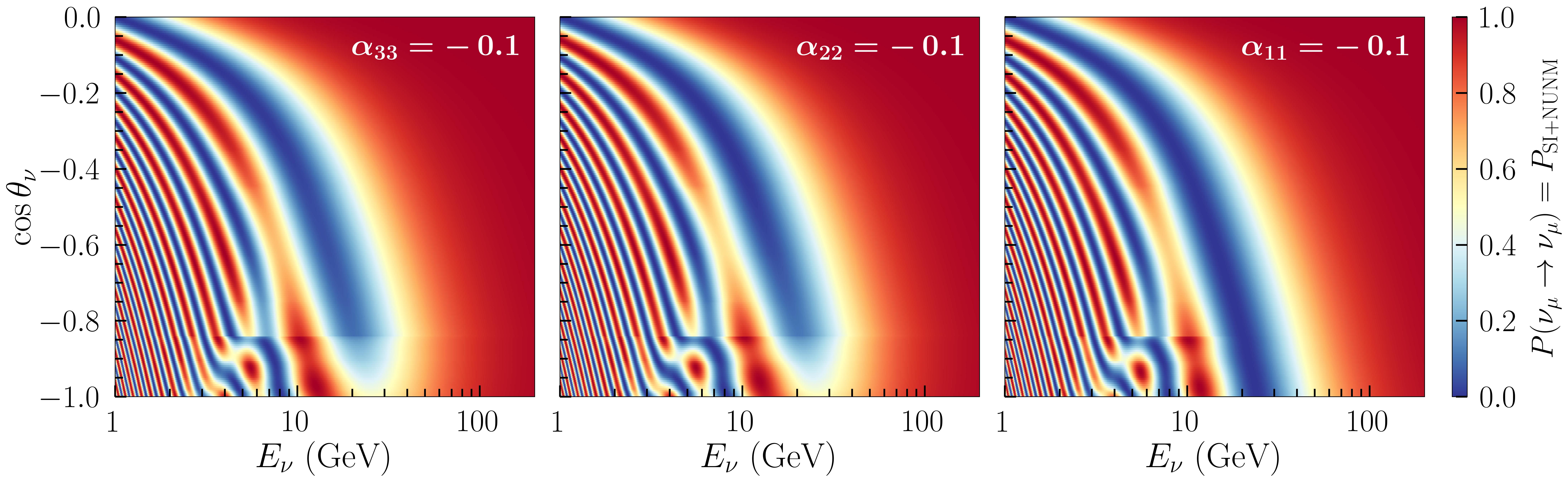}
	\caption{Neutrino oscillation probabilities corresponding to the disappearance channel, $ P\,(\nu_\mu \rightarrow \nu_\mu)$, shown in the plane of $E_\nu$ and $\cos \theta_\nu$ for the SI $+$ NUNM scenario. The left, middle, and right panels correspond to the diagonal NUNM parameters $\alpha_{33}$, $\alpha_{22}$, and $\alpha_{11}$, respectively. We take $\alpha_{ii}\,=\,-\,0.1$, considering one nonzero NUNM parameter at a time.}
	\label{fig:osc_diag_neg}
\end{figure*}

\begin{figure*}[htp!]
	\centering
	\includegraphics[width=\textwidth]{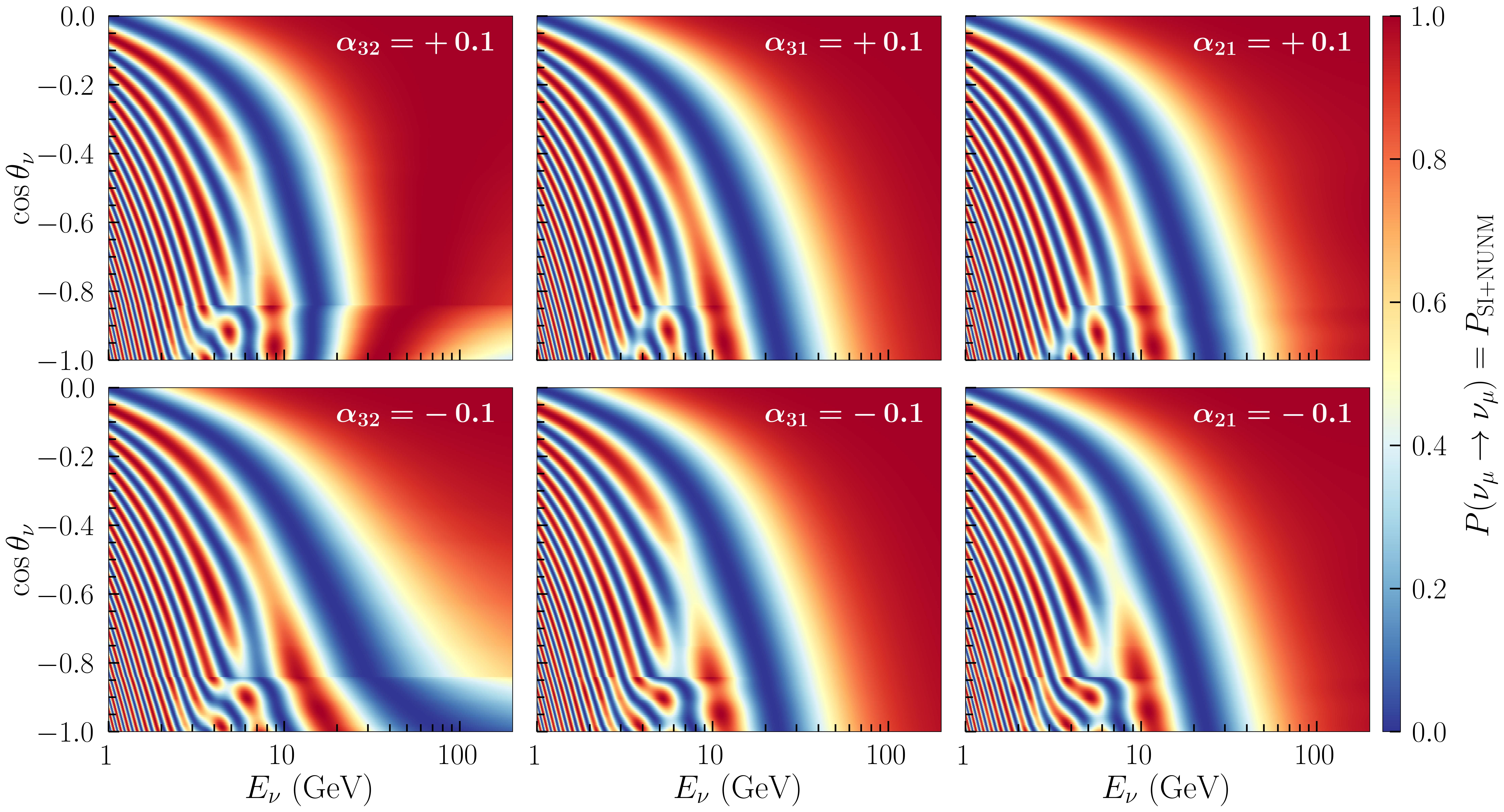}
	\caption{Neutrino oscillation probabilities corresponding to the disappearance channel, $ P\,(\nu_\mu \rightarrow \nu_\mu)$, shown in the plane of $E_\nu$ and $\cos \theta_\nu$ for the SI $+$ NUNM scenario. The left, middle, and right panels correspond to the off-diagonal NUNM parameters $\alpha_{32}$, $\alpha_{31}$, and $\alpha_{21}$, respectively. The top (bottom) panel shows the probability for a positive (negative) value of 0.1 ($-0.1$) of the NUNM parameters, $\alpha_{ij}$. We consider one nonzero NUNM parameter at a time.}
	\label{fig:osc_offdiag_mumu}
\end{figure*}

This appendix presents the neutrino oscillation probabilities under the SI $+$ NUNM hypothesis, for nonzero values of the NUNM parameters taken one at a time. Figure~\ref{fig:osc_diag_neg} presents the oscillation probability $P(\nu_\mu \rightarrow \nu_\mu)$ for the diagonal NUNM parameters with $\alpha_{ii} = -\,0.1$ in the plane of $E_\nu$ and $\cos \theta_\nu$. The left, middle, and right panels correspond to $\alpha_{33}$, $\alpha_{22}$, and $\alpha_{11}$, respectively. As discussed earlier, the diagonal parameters $\alpha_{33}$ and $\alpha_{22}$ modify $\theta_{23}$, thereby affecting the depth of the oscillation valley, while $\alpha_{11}$ modifies the effective matter potential through its coupling to matter effects.

Figure~\ref{fig:osc_offdiag_mumu} presents the oscillation probability $P(\nu_\mu \rightarrow \nu_\mu)$ for the off-diagonal NUNM parameters in the plane of $E_\nu$ and $\cos \theta_\nu$. The top and bottom panels correspond to $\alpha_{ij} = +0.1$ and $\alpha_{ij} = -0.1$, respectively. The left, middle, and right panels show the oscillation probabilities for $\alpha_{32}$, $\alpha_{31}$, and $\alpha_{21}$, respectively. The off-diagonal parameter $\alpha_{32}$ causes the oscillation valley to bend, with the bending occurring inward (outward) for positive (negative) values of the parameter. The parameters $\alpha_{31}$ and $\alpha_{21}$ alter the matter effects experienced by neutrinos; the matter effects are suppressed for positive values of the parameters and enhanced for negative values. Consequently, this results in an increase (a decrease) in $P(\nu_\mu \rightarrow \nu_\mu)$ for $\alpha_{ij} > 0$ ($\alpha_{ij} < 0$) in the region of matter effect.

\section{Systematic parameters at best-fit values}
\label{app:bestfit}

Our analysis incorporates uncertainties associated with various systematics, including oscillation parameters, atmospheric neutrino flux, cross section, detector effects, and normalization. The best-fit values of these nuisance parameters obtained during the fit for the NUNM parameters are summarized in Table~\ref{tab:combined_nuisance_parameters}. The second to sixth columns present the best-fit values of the systematic parameters obtained from the fit to data under the SI $+$ NUNM hypothesis, with one NUNM parameter fitted at a time; while the last two columns present the nominal values and the $1\sigma$ priors (if available) of these nuisance parameters. For parameters with available prior values, a pull penalty has been applied to the $\chi^2_\text{mod}$. For parameters with no prior, we mention ``\textit{Unconstrained}'', indicating that no pull penalty is applied for these parameters during the fit.

\begin{table*}[t]
	\centering
	\small
	\setlength{\tabcolsep}{5pt}
	\renewcommand{\arraystretch}{1.2}
	
	\begin{tabular}{lcc@{\hspace{20pt}}ccccc}
		\hline\hline
		\multirow{3}{*}{\textbf{Parameters}}
		& \multicolumn{5}{c}{\textbf{Best-fit values}}
		& \multirow{3}{*}{\textbf{Nominal value}}
		& \multirow{3}{*}{\textbf{Prior ($1\sigma$)}} \\
		\cmidrule(lr){2-6}
		& \multicolumn{2}{c}{\textbf{Diagonal}}
		& \multicolumn{3}{c}{\textbf{Off-diagonal}} \\
		\cmidrule(r{10pt}){2-3}
		\cmidrule(l{10pt}){4-6}
		& $\alpha_{22}$ & $\alpha_{33}$
		& $\alpha_{21}$ & $\alpha_{31}$ & $\alpha_{32}$
		& & \\
		\hline
		\multicolumn{8}{l}{\textbf{Detector:}} \\
		DOM efficiency     & 1.064 & 1.064 & 1.061 & 1.064 & 1.065 & 1.0 & $\pm\,0.1$ \\
		Ice absorption     & 0.974 & 0.974 & 0.974 & 0.974 & 0.974 & 1.0 & Unconstrained \\
		Ice scattering     & 0.987 & 0.987 & 0.989 & 0.989 & 0.988 & 1.05 & Unconstrained \\
		Relative eff. $p_0$ & $-0.269$ & $-0.269$ & $-0.257$ & $-0.265$ & $-0.272$ & 0.10 & Unconstrained \\
		Relative eff. $p_1$ & $-0.043$ & $-0.043$ & $-0.041$ & $-0.044$ & $-0.043$ & $-0.05$ & Unconstrained \\
		\midrule
		\multicolumn{8}{l}{\textbf{Atmospheric neutrino flux:}} \\
		$\Delta\gamma_\nu$ & 0.064 & 0.064 & 0.068 & 0.064 & 0.064 & 0.0 & $\pm\,0.1$ \\
		$\Delta\pi^+$ yields [A--F] & 0.061 & 0.061 & 0.058 & 0.058 & 0.061 & 0.0 & $\pm\,0.3$ \\
		$\Delta\pi^+$ yields G & $-0.055$ & $-0.055$ & $-0.073$ & $-0.059$ & $-0.048$ & 0.0 & $\pm\,0.3$ \\
		$\Delta\pi^+$ yields H & $-0.018$ & $-0.018$ & $-0.029$ & $-0.019$ & $-0.013$ & 0.0 & $\pm\,0.15$ \\
		$\Delta K^+$ yields W & 0.085 & 0.085 & 0.099 & 0.087 & 0.077 & 0.0 & $\pm\,0.4$ \\
		$\Delta K^+$ yields Y & 0.107 & 0.107 & 0.132 & 0.109 & 0.094 & 0.0 & $\pm\,0.3$ \\
		$\Delta K^-$ yields W & $-0.009$ & $-0.009$ & $-0.0004$ & $-0.010$ & $-0.013$ & 0.0 & $\pm\,0.4$ \\
		\midrule
		\multicolumn{8}{l}{\textbf{Neutrino interaction cross section:}} \\
		$M_A^{\mathrm{CCQE}}$ (in $\sigma$)
		& 0.062 & 0.062 & 0.071 & 0.086 & 0.057 & 0.0 & $\pm\,1.0$ \\
		$M_A^{\mathrm{CCRES}}$ (in $\sigma$)
		& 0.607 & 0.606 & 0.598 & 0.634 & 0.599 & 0.0 & $\pm\,1.0$ \\
		DIS CSMS
		& 0.034 & 0.034 & 0.038 & 0.065 & 0.045 & 0.0 & $\pm\,1.0$ \\
		$\sigma_{\mathrm{NC}}/\sigma_{\mathrm{CC}}$
		& 1.127 & 1.127 & 1.103 & 1.113 & 1.129 & 1.0 & $\pm\,0.2$ \\
		\midrule
		\multicolumn{8}{l}{\textbf{Normalization:}} \\
		$A_{\mathrm{eff}}$ scale
		& 0.824 & 0.824 & 0.826 & 0.825 & 0.823 & 1.0 & Unconstrained \\
		\midrule
		\multicolumn{8}{l}{\textbf{Atmospheric muons:}} \\
		Atm.\ $\mu$ scale
		& 1.365 & 1.365 & 1.291 & 1.318 & 1.348 & 1.0 & Unconstrained \\
		\midrule
		\multicolumn{8}{l}{\textbf{Oscillations:}} \\
		$\theta_{23}$
		& $45.358^\circ$ & $45.327^\circ$
		& $46.254^\circ$ & $44.138^\circ$ & $45.342^\circ$
		& $45.573^\circ$ & Unconstrained \\
		$\Delta m^2_{31}$ (eV$^2$)
		& 0.002489 & 0.002489
		& 0.002540 & 0.002570 & 0.002499
		& 0.002484 & Unconstrained \\
		\hline\hline
	\end{tabular}
	\caption{The best-fit values of the systematic parameters obtained from fits to the data sample for the diagonal NUNM parameters, $\alpha_{22}$ and $\alpha_{33}$ (second and third columns, respectively) and the off-diagonal NUNM parameters, $\alpha_{21}$, $\alpha_{31}$, and $\alpha_{32}$ (fourth, fifth, and sixth columns, respectively). During the fit, we consider one NUNM parameter at a time. The nominal values and the corresponding $1\sigma$ priors (if available) of the nuisance parameters are shown in the seventh and eighth columns, respectively.}
	\label{tab:combined_nuisance_parameters}
\end{table*}

\section{Data-MC comparison}
\label{app:data_MC}

\begin{figure}[htp!]
	\centering
	\includegraphics[width=\linewidth]{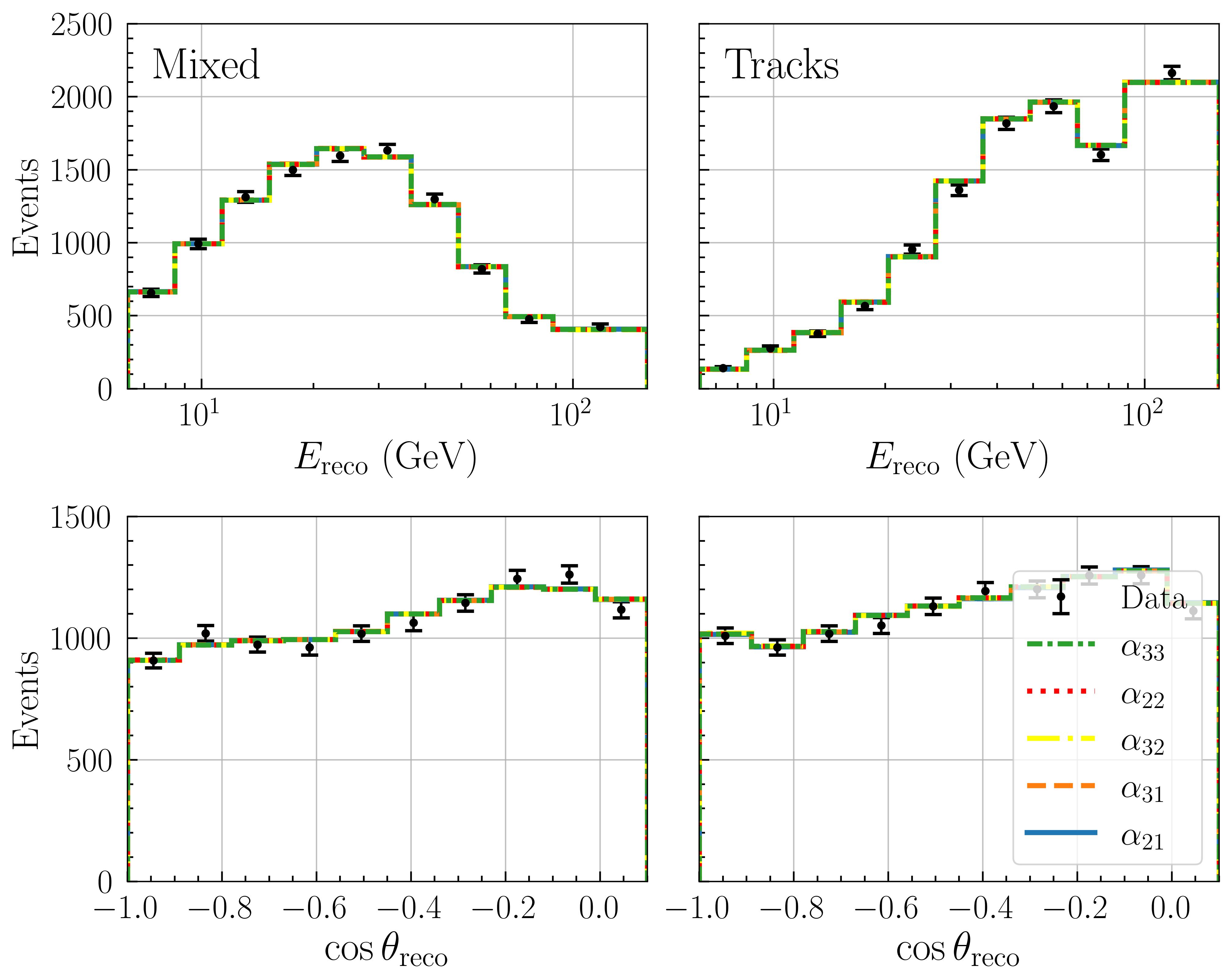}
	\caption{Comparison of the observed data and the best-fit MC events as a function of reconstructed energy, $E_\text{reco}$ (top panel), and reconstructed cosine of zenith angle, $\cos \theta_\text{reco}$ (bottom panel). The left (right) panels stand for the mixed (tracklike) event topology. The black dots represent the observed event counts from the 8-year golden sample of IceCube DeepCore, with the bars representing their statistical uncertainties. The MC event distribution for the NUNM parameters are shown by the colored histograms, which overlap with each other.}
	\label{fig:data_mc}
\end{figure}

In this section, we present a comparison between the observed data from the 8-year golden event sample of IceCube DeepCore and the expected MC events generated using the best-fit values obtained under the SI $+$ NUNM hypothesis. Figure~\ref{fig:data_mc} shows the 1D projections of the observed data and the corresponding MC events in the reconstructed energy (top row) and cosine of zenith angle (bottom row). The left and right panels correspond to mixed and tracklike event topologies, respectively. The observed data are shown by the black dots with error bars. The green, red, yellow, orange, and blue histograms correspond to the expected MC event distributions obtained using the best-fit values of $\alpha_{33}$, $\alpha_{22}$, $\alpha_{32}$, $\alpha_{31}$, and $\alpha_{21}$, respectively, considering one NUNM parameter at a time. The MC distributions thus obtained are found to be consistent with the observed data in all bins.

\bibliographystyle{apsrev4-2}
\bibliography{deepcore_nunm_ref}

\end{document}